\documentclass{aa}
\usepackage[varg]{txfonts}
\bibpunct{(}{)}{;}{a}{}{,} 
\usepackage{graphicx}
\usepackage[breaklinks, colorlinks, linkcolor=blue, citecolor=blue]{hyperref} 
\usepackage{booktabs}
\usepackage{amsmath}
\usepackage{ulem}
\usepackage{bm}

\makeatletter
\renewcommand*\aa@pageof{, page \thepage{} of \pageref*{LastPage}}
\makeatother

\begin{document}

\title{The IAG spectral atlas of the spatially resolved Sun: Centre-to-limb observations \thanks{Data and electronic version of atlas: \url{ https://www.astro.physik.uni-goettingen.de/research/solar-lib}} }

\author{M. Ellwarth
  \and S. Sch\"afer
  \and A. Reiners
  \and M. Zechmeister}

\institute{Institut f\"ur Astrophysik und Geophysik, Georg-August Universit\"at G\"ottingen, Friedrich-Hund-Platz 1, 37077 G\"ottingen, Germany}

\date{\today}

\abstract {Solar surface magneto-convection appears as granulation pattern that impacts spectral lines in terms of both shape and wavelength. Such induced effects also tend to vary  over the observed solar disc because of the changing observation angle and, thus, the changing observation height as well. Centre-to-limb observations of the resolved Sun offer an insight into the variable spectral behaviour across different heliocentric observing positions, providing crucial information about limb darkening, convective velocities, and line profile variability relevant to radial velocity (RV) calculations. Thus, RV measurements and exoplanet transit spectroscopy depend on precise reference templates.}
{We want to provide a spectroscopic centre-to-limb solar atlas at high spectral resolution and high-frequency accuracy. The atlas shall help improve the understanding of the solar atmosphere and convection processes.}
{We performed high-resolution observations of the resolved quiet Sun with a Fourier transform spectrograph at the Institut f\"ur Astrophysik und Geophysik in G\"ottingen. Our dataset contains a wavelength range from 4200\,\r{A} to 8000\,\r{A}. We obtained 165 spectra in total, with a spectral resolution of $\Delta \nu$ = 0.024\,cm$^{-1}$, corresponding to a resolving power $R$ of 700,000 at $\sim$6000\,\r{A}.} {We present a centre-to-limb solar atlas containing 14 heliocentric positions. To check for consistency, we investigated the \ion{Fe}{I}~6175\,\r{A} absorption line and compared our line profiles with previous centre-to-limb observations and also with simulations. The line profile and also the bisector profiles are generally consistent with previous observations, but we have identified\ differences to model line profiles, especially close to the solar limb.}{}


\titlerunning{The IAG spectral atlas of the spatially resolved Sun}
\maketitle

\section{Introduction}

The solar surface is covered in a pattern of granulation. Plasma rises up in convection cells and after cooling, it sinks down at the rims of the cells again \citep[e.g.][]{1967sogr.book.....B}. This convection pattern impacts the appearance of the solar spectrum. Because of the interplay of granules and intergranular lanes, spectral lines change shape and wavelength. Observations covering areas with multiple granules end up with an intrinsic radial velocity (RV) shift. The blue-shifted uprising granules, which are brighter and hotter, contribute more light to the line profile than their  cooler, downwards-moving counterparts of red-shifted intergranular lanes \citep[e.g.][]{2004suin.book.....S}. The motion of these dynamic cells ends up as a net convective blue-shift for most photospheric absorption lines \citep{1978SoPh...58..243B} and depends on line depth, excitation potential, and wavelength region \cite[][]{1981A&A....96..345D}. Because of the decrease of granular contrast towards the solar limb, convective blue-shift decreases from disc centre to limb.

At the beginning of the 20th century, \cite{1907AN....173..273H} had already discovered that absorption lines observed on the limb of the solar disc show a shift to longer wavelengths compared to the ones at the disc centre.  Because the Sun is active and convection is not a stable process, these RV shifts also contain variations over time. Additionally to the wavelength shifts from centre-to-limb, photospheric lines show a decrease in line depth towards the limb \cite[see e.g.][]{2018A&A...611A...4L}. The effect is primarily caused by limb darkening originating from the temperature decrease towards higher atmospheric layers. Additionally, non-LTE effects as well as the formation height vary across line profiles. Therefore, resolved Sun spectra offer an insight into the spectral changes of limb darkening, helping to improve our picture of the atmospheric structure of our home star.

Because of its immediate proximity to Earth, the Sun is the only star we can observe in such great detail and, for example, with the aim of resolving its surface. Therefore, it serves as a perfect benchmark star to investigate these variabilities over time and centre-to-limb relations. Precise and highly resolved surveying and correction models of the solar behaviour are essential in many fields in solar and stellar physics \citep{2016A&A...595A..26R, 2021A&A...649A..17D, 2022AJ....163...11P}. Centre-to-limb observations help to create more precise templates of stars and therefore accurate measurements of the Sun are required to support the detection of exoplanets \citep{2021arXiv210714291C}. For example, since the background star for exoplanet hunting is generally not spatially resolvable, it is crucial to work with a precise template to subtract the correct stellar spectrum. \cite{2017A&A...605A..90D} used synthetic 3D hydrodynamic photospheric line profiles for stellar atmospheres to mimic an exoplanet transit with the aim of identifying features that seem most realistic to measure in future observations. For this kind of study, it is mandatory to have reliable and detailed templates.

Another important issue in stellar physics is the intrinsic variability of stars in general. This variability is currently the most significant obstacle in terms of extreme precision radial velocity (EPRV) measurements. The desired precision of the EPRV\ is at a level of less than 10~cm\,s$^{-1}$ and is thus an essential technique on the path to finding Earth analogues. The report from \cite{2021arXiv210714291C} emphasised the key importance of expanding our knowledge on solar and stellar variability resulting from surface phenomena. This is because these effects are on the order of tens of m/s and, thus, considerably higher than the desired signal of a second Earth.

Furthermore, spatially resolved solar spectra are an important tool in validating and improving the accuracy of atmospheric models of the Sun. \cite{2009A&A...507..417P} tested 3D models against high-spatial-resolution observations of different absorption lines, especially focusing on oxygen. In another article, \cite{2009A&A...508.1403P} focused on NLTE line formation, and solar oxygen abundance. Overall the authors found a high consistency between 3D models and observations, particularly for oxygen lines. For comparison of 3D magnetohydrodynamic (MHD) simulations and 1D averaged atmospheres, \cite{2011ApJ...736...69U} investigated the behaviour of the \ion{Fe}{I} 5250~\r{A} and 5253~\r{A} and \ion{CO}{} 46636~\r{A} lines. One of the authors' main findings was the broad divergence between the spectral lines of the different simulations. This divergence in shape and intensity becomes especially stronger  towards the solar limb, pointing out the inconsistency among simulations. \cite{2021ApJ...923..207C} compared the 3D radiative-magnetohydrodynamics simulation codes MURaM and STAGGER, and found small differences in the synthetic spectra that dissolve by reducing the resolution to an observational level. In their work, they concluded that future diagnostics efforts of comparing different 3D models against each other and also comparing them to centre-to-limb observations.

There are already several spatially resolved observations and investigations of the Sun. \cite{1988A&AS...72..473B} published parameters of 143 solar absorption lines for different positions of the solar disc and thereby provided one of the first detailed listings of centre-to-limb variations. \cite{2015A&A...573A..74S} compared the spectra of the solar limb with those of the disc centre. The author constructed an atlas of the limb/disc ratio to clarify the variation between these two intensity spectra. 
\cite{2019KPCB...35...85O} determined bisectors from the spectral lines of the resolved Sun. For their analysis, they used Fraunhofer lines located in the wavelength regions around 5320~\r{A} and 5390~\r{A} and showed the quantitative changes of the lines from centre to limb. In a larger observation campaign, \cite{2018A&A...611A...4L,2019A&A...622A..34S,2019A&A...624A..57L} observed the Sun at ten heliocentric positions and an accuracy at m/s for limited wavelength ranges in the visible. The authors examined 11 solar lines to investigate centre-to-limb variation, the convective blue-shift, and the impact of spectral resolution on these velocity shifts. They confirmed that the convective blue-shift varies across the solar disc due to the different heliocentric observation positions and that the effect is stronger for weaker lines. \cite{2022AJ....163...11P} used the observed FeI 5434.5~\r{A} line of \cite{2018A&A...611A...4L} to create a synthetic disk-integrated spectrum. Further investigations of the centre-to-limb variation of line profile properties and comparisons of those values against models have been undertaken \cite[e.g.][]{2004A&A...423.1109A,2017PASJ...69...46T,2019SoPh..294...63T,2022SoPh..297....4T}, which help to evaluate the fidelity of stellar and solar models.

These previous studies show that there is a significant interest and need for resolved Sun spectra. Here, we present a novel ultra-high-resolution spectral atlas with a wavelength range between 4200\,\r{A} and 8000\,\r{A} which will improve our understanding of the solar photosphere as well as that of stars overall. 

In this paper, we first describe our instrumental set-up in Sect.~\ref{Inst} and the observations we carried out during this observation campaign in Sect.~\ref{Obs}. In Sect.~\ref{Reduction}, we give an overview of the applied data calibration and discuss intrinsic error sources of the data set. Finally, we present the FeI 6175~\r{A} solar absorption line of our spatially resolved solar surface atlas and compare it with former observations as well as with simulations in Sect.~\ref{Results}. A first insight into our data was already given in \cite{2021MNRAS.tmp.1964B}, where they used the atlas in comparison to 1D LTE, 1D NLTE, and 3D NLTE line profiles for a reanalysis of the solar oxygen abundance.

\section{Instruments} \label{Inst}
\begin{figure}
        \centering
        \includegraphics[width=0.5\textwidth]{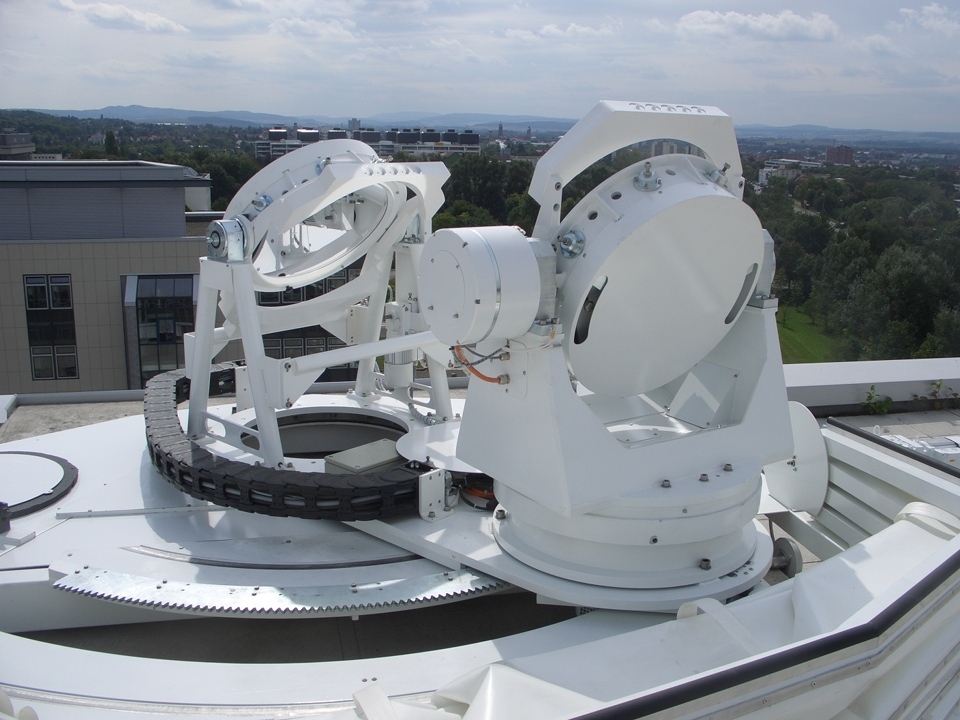}
        \caption{Siderostat of the VVT on the rooftop at the IAG in G\"ottingen. The window to the vacuum tube of the telescope can be seen below the mirror on the left hand side.}
        \label{fig:telesc}
\end{figure}

At the Institut für Astrophysik und Geophysik (IAG) in G\"ottingen, we carried out solar observations with the Vacuum Vertical Telescope (VVT) that is integrated into the faculty building. The corresponding siderostat is mounted at the rooftop of the building (see Fig.~\ref{fig:telesc}). 
In combination with a Fourier transform spectrograph (FTS) and our resolved Sun set-up, we observed spatially resolved areas on the solar surface across a broad spectral range. In this section, we present a short introduction to the set-up and the FTS.

\begin{figure}
        \centering
        \includegraphics[width=0.5\textwidth]{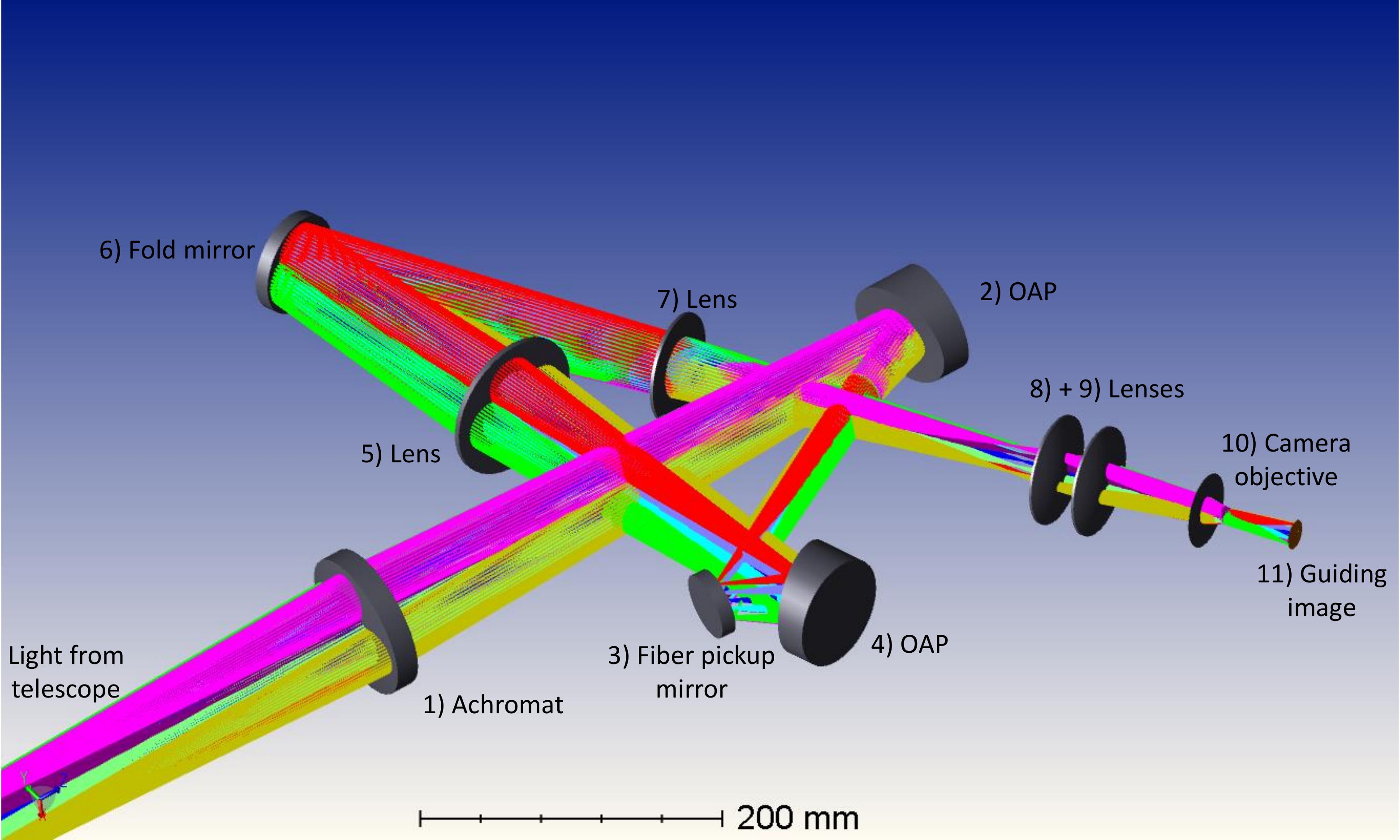}
        \caption{Schematic resolved Sun set-up. The sunlight gets first collimated by an achromat lens (1). An off-axis parabolic (OAP) mirror (2) is re-imaging the solar disc on the fibre pickup mirror (3). The remaining light is send to another OAP (4), through a lens (5), gets folded at (6), and after three more lenses (7-9) it ends on the guiding camera.}
        \label{fig:setup}
\end{figure}

\subsection{Resolved Sun set-up}

\begin{figure}
        \centering
        \includegraphics[width=0.45\textwidth]{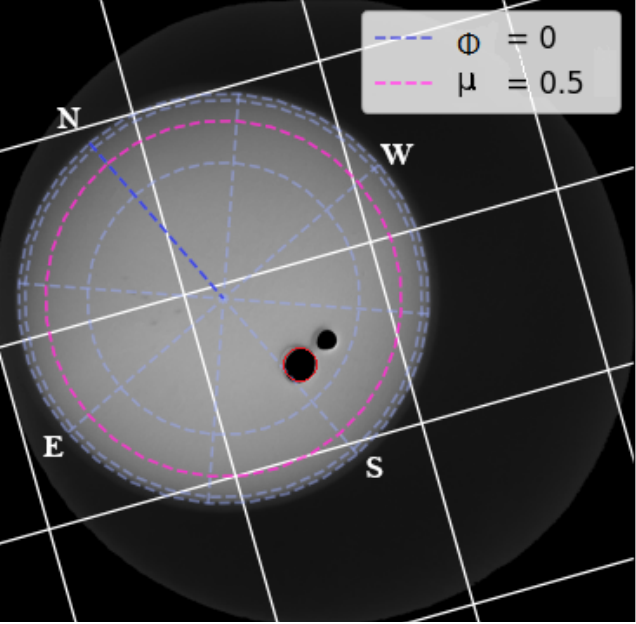}
        \caption{Image of the solar disc on the fibre pick-up mirror. The dashed lines on the solar image represent different observation angles respectively directions. The larger hole marked in red shows the fibre port. The smaller black spot originates by a hole in a mirror that is located further up in the telescope.}
        \label{fig:sun}
\end{figure}

The optical set-up is shown schematically in Fig.~\ref{fig:setup}. Specifically, the light of the whole solar disc gets directed from the telescope into the set-up. On the image plane of the optical set-up, the fibre pickup mirror~(3) is mounted, which has a hole with an integrated hexagonal-shaped fibre (diameter of 525\,\textmu m), where the desired spatially resolved sunlight is collected. In Fig. \ref{fig:sun}, the whole solar disc is displayed on the fibre pickup mirror with a solar coordinate system plotted on top. The larger, red tagged circle on the disc marks the fibre port. The fibre itself is about one-fifth the size of the fibre port and is not visible in Fig.~\ref{fig:sun}. This fibre position is used to pick the observing area on the solar surface. By changing the position of the solar disc on the fibre pickup mirror, we can choose the required solar coordinates. With this set-up, the field of view of an observation covers 32.5\arcsec ($\approx$ 23.600~km) in diameter on the solar disc. For more detailed information about our resolved Sun set-up, we refer to \cite{10.1117/12.2560156}.

\subsection{Fourier transform spectrograph}
The light from the fibre leads to a Bruker IFS 125 HR FTS, a spectrograph based on the Michelson-Interferometer. Such instruments divide the incoming light with a beam-splitter, which deflects one part of the light to a fixed mirror and the other to a movable mirror. The reflected light recombines and interferes at the beam-splitter and gets (partly) directed to the detector. 
The resulting intensity function over the position of the movable mirror creates an interferogram. Performing a fast Fourier transformation on this interferogram recovers the spectral information of the light. The maximal spectral resolving power of our FTS is about $R \approx 10^6$. The set-up works with a Quartz beam-splitter and a Si detector, leading to a general spectral range between 4050\,\r{A} and 10650\,\r{A}.
We used a dichroic filter to cut all wavelengths after 8000~\r{A}, since our initial set-up was designed to use the wavelength range after 8000~\r{A} for drift measurements with a Fabry-Pérot. However, during our observations, this feature was not fully implemented. After half of the observation campaign, we removed the filter because it introduced small absorption gaps to our spectra (see also Sect. \ref{norm}).

\section{Observations} \label{Obs}
\begin{figure}[ht]
        \centering
        \includegraphics[width=0.5\textwidth]{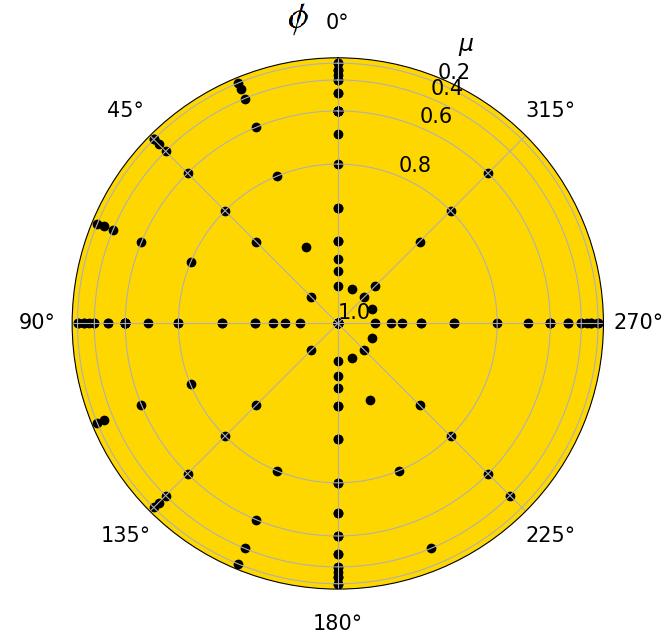}
        \caption{Overview of all observed positions on the solar surface. Size of the dots corresponds to the fibre size of the observations.}
        \label{fig:sun_obs}
\end{figure}

For the observation campaign of this work we used the double-sided mode of the FTS and chose a spectral resolution of $\Delta \nu$ = 0.024~cm$^{-1}$ corresponding to a resolving power of $R = \nu/\Delta \nu \approx$ 700,000 at $\lambda$\,=\,6000\,\r{A}. The spectral range we receive in the end is between 4200\,\r{A} and 8000\,\r{A}. The observations were carried out between April 2020 and June 2021 close to a solar minimum. By including 165 individual observations, the predominant focus of this observing campaign was to resolve the centre-to-limb variations spectrally. To neglect surface effects such as supergranulation, we decided to observe positions all around the solar disc and eventually average over the respective heliocentric radii.
Our observations include 14 different heliocentric observation radii, which are defined as
\begin{align}
    \mu = \cos \theta.
\end{align}
The heliocentric angle $\theta$ constitutes the angle between the perpendicular line to the solar surface and the line of sight from Earth. That implies a value of $\mu = 1$ for the disc centre and a non-linearly decreasing $\mu$ value the farther to the solar limb one is looking with $\mu = 0$ for the limb itself. The observations were carried out at different $\phi$ angles on the Sun, where $\phi$=0 is defined to direct to the solar north pole. In Fig. \ref{fig:sun_obs}, all disc positions observed for this campaign are marked, while Table~\ref{tab:values} lists the positions of our observations together with the total observing time of each $\mu$ and the corresponding data characteristics. For this data set, we tried to ensure that all observed areas contain only quiet Sun regions. This excludes magnetic active regions like sunspots or plage. Therefore we checked the observing areas for disturbances by looking at the daily solar images of the Solar Dynamics Observatory (SDO) and live images from our telescope.
However, there is still a lot of motion remaining even in quiet Sun regions. To obtain the most unbiased spectra, the goal is to diminish these velocity effects as much as possible. Important sources of internal solar noise are $p$-mode oscillations of acoustic waves in the interior of the Sun, which lead to oscillations on the surface. These oscillations have predominantly periods between three and eight minutes with the largest amplitude in the photosphere of approximately five~minutes and a typical velocity amplitude of 0.5 to 1~km\,s$^{-1}$. These so-called five-minute oscillations, propagate predominantly vertically to the solar surface \citep{1999A&A...346..633S}, and therefore their velocity impact become less pronounced, the farther observations are taking place to the solar limb. Since spatial areas of about 30~Mm in diameter on the Sun (resp. $\approx$ 40\arcsec) often oscillate in phase \citep{2004suin.book.....S} and our fibre observes a smaller area, we decided to average over this five-minute oscillation by observing a multiple of this time span. We chose a total observation time of about ten minutes per spectrum. This includes ten cycles of the movable FTS mirror (about one minute each cycle).\\

\begin{table}
        \centering
        \small
        \caption{Observed $\mu$-positions.}
        \begin{tabular}{@{}lrrrrrr@{}}
                \hline
                \hline
             & Obs. & &  &  &  & \\ 
                $\mu$ & time & S/N & $\Delta v_{\rm point.}$ & $\sigma_{\rm scat}$ & $\Delta v_{\rm rot}$ & $\Delta v_{{\rm LARS},6175}$\\
                & [min] & & [m/s] & [m/s] & [m/s] & [m/s]\\
                \midrule
                1.00   & 170 & 640 & 4.3 & 0.6 &- & 15.9\\
                0.99 & 120 & 500 & 9.2 & 2.4 & 0.55 &-\\
                0.98  & 50 & 500 & 9.5 & 9.3 & 0.36 & -\\
                0.97 & 40 & 690 & 10.0 & 12.1 & 0.28& -\\
                0.95 & 100 & 590 & 10.8 & 6.2 & 0.18&48.6\\
                0.90 & 120 & 690 & 12.5 & 9.4 & 0.08&10.2\\
                0.80 & 150 & 580 & 15.1 & 0.02 & 0.04&23.5\\
                0.70 & 50 & 550 & 17.1 & 18.9 & 0.16&31.4\\
                0.60 & 310 & 570 & 18.6 & 4.2 & 0.30&12.7\\
                0.50 & 80 & 580 & 19.9 & 0.06 & 0.52&26.5\\
                0.40 & 150 & 530 & 20.8 & 8.4 & 0.89&3.6\\
                0.35 & 40 & 610 & 21.1 & 5.1 & 1.22&-\\
                0.30 & 120 & 510 & 21.5 & 12.4 & 1.71&8.8\\
                0.20 & 150 & 420 & 22.1 & 1.9 & 4.02&-\\
                \hline
        \end{tabular}
        \tablefoot{Listed: $\mu$ positions, the total observing time of all spectra of the corresponding $\mu$ (e.g. 17 single spectra add up to 170\,min of observing time), and the S/N of single spectra. Errors: Maximum pointing position offset $\Delta v_{point.}$, velocity dispersion among spectra $\sigma_{scat}$, the maximum rotational velocity offset $\Delta v_{rot}$, and  the line accuracy $\Delta v_{{\rm LARS},6175}$.}
\label{tab:values}
\end{table}             

Another important noise source is supergranulation. Different from $p$-modes, supergranules expand predominantly horizontally and therefore have a stronger impact on the total velocity flow depending on how close the observation is  carried out with respect to the solar limb. Supergranulation flows usually show velocities in the range of 300\,m\,s$^{-1}$ to 500\,m\,s$^{-1}$ but can also go up to more than 1~km\,s$^{-1}$ \citep{2000SoPh..193..313S}. The size of these supergranules is in the same order of around 30~Mm as the vertical oscillations. Since supergranulation has oscillation periods of up to a day, it is impossible to average over it by simply adjusting our scanning time. By observing different $\phi$ angles, we are ultimately able to reduce supergranulation effects  by averaging the spectra.

\section{Data} \label{Reduction}
In this section, we go through the applied data processing methods and outline the reliability of our data set. We outline different error sources emerging and examine the illumination of our pick-up fibre for all heliocentric $\mu$ radii to understand the extent of the superpositions of the $\mu$ ranges towards the limb.

\subsection{Data calibration}
As mentioned in the last two sections, our data was observed using an FTS. These kinds of spectrographs provide a couple of advantages compared to other spectrographs. Besides a high spectral resolution, it is possible to scan wavelength ranges over thousands of angstroms simultaneously during a short timeframe. A detailed account on Fourier transform spectroscopy can be found, for example, in \cite{2001ftsp.book.....D} or \cite{2007ftis.book.....G}. Nevertheless, the following sub-sections provide a brief overview about FTS data processing we applied for this research. 

\subsubsection{Wavenumber correction}

The FTS is equipped with a He-Ne reference laser as a measuring scale for mirror retardation. The laser beam follows, in theory, a  path parallel to the science light. However, in practice, the laser beam is usually slightly tilted to the beam of the light source. This circumstance and the fact that the divergence of the laser beam itself is always smaller than the one of the science light introduces a wavenumber offset between the science and calibration path. Accordingly, a wavenumber correction is necessary.
A detailed description of this technical background can for example be found in Chapter 2 of \cite{2007ftis.book.....G}. The intrinsic wavenumber offset of an FTS is linear in wavenumber \citep{2004JOSAB..21.1543S}. Therefore, the precision of the wavelength solution is stable with only one free parameter for wavelength correction. To correct the intrinsic wavenumber offset of the FTS, we calculate the correction factor 
\begin{align}
    \kappa \approx -\frac{\Delta \nu}{\nu_0}.
\end{align}
The $\kappa$ factor is a constant parameter for the whole wavenumber range, $\nu,$ and gets calculated by the shift, $\Delta \nu,$ between a known reference wavenumber, $\nu_0$, and the corresponding observed wavenumber. Eventually, the wavenumber solution of the FTS gets corrected by
\begin{align}
        \nu = \nu_{\rm FTS}~(1+\kappa),
\end{align}
where $\nu_{\rm FTS}$ is the original and $\nu$ the corrected wavenumber solution. A common method in spectroscopic calibration is the use of telluric absorption lines of Earth's atmosphere as reference wavelengths. For the reference line position $\nu_0$ we use O$_2$ lines of the HITRAN2020 database \citep{GORDON2022107949} and determined the line shifts, $\Delta \nu,$ between HITRAN2020 wavenumbers and the observed tellurics in the solar spectra. We calculated 100 O$_2$ line shifts for every single spectrum to ultimately compute $\kappa$.
\begin{figure*}
        \centering
        \includegraphics[width=\textwidth]{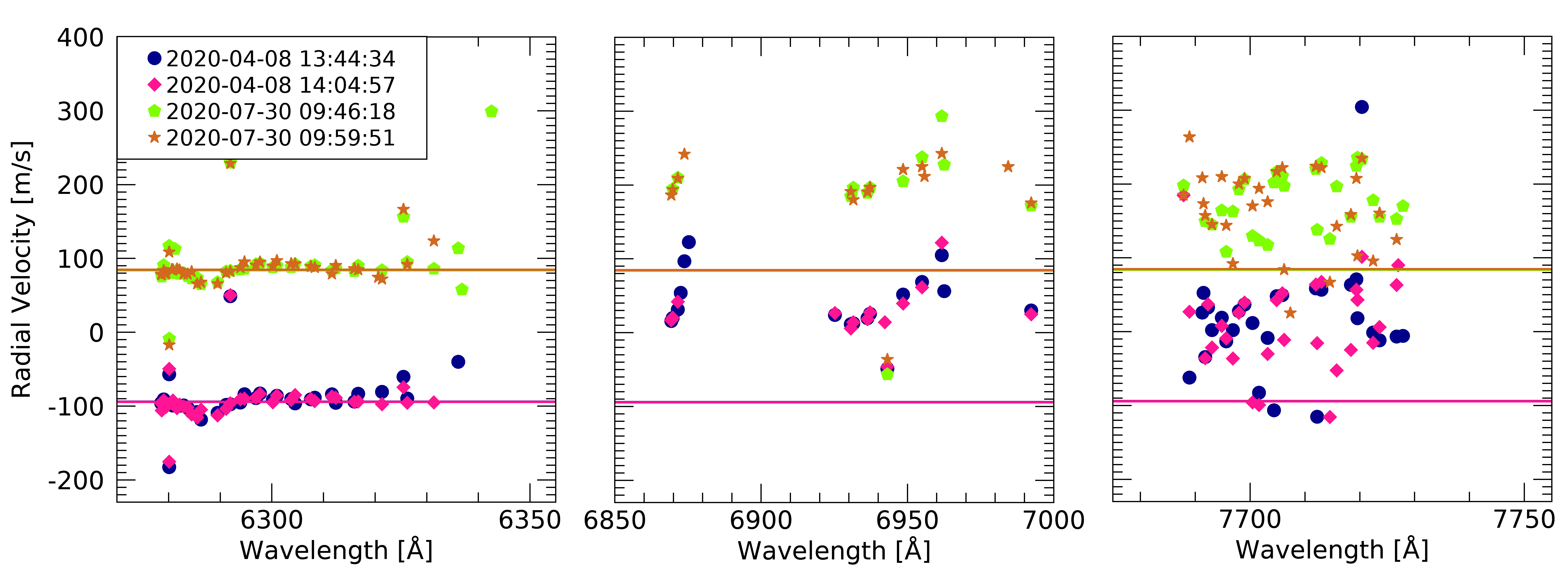}
        \caption{Calculated radial velocities, $v = \Delta \nu/\nu_0 \cdot c,$ of line shifts between about 100 observed telluric lines and the reference wavelength of the HITRAN2020 line list. The panels show $v$ values for three different wavelength ranges (vacuum wavelength) and four spectra of two different months. The horizontal lines mark the respective median of the velocities of the first panel.}
        \label{fig:kfac}
\end{figure*}

Figure \ref{fig:kfac} shows the velocities $v = \Delta \nu/\nu_0 \cdot c$, which correspond to $\kappa$ for four different observations. The plot shows the velocities between the observed O$_2$ lines and the reference wavelengths of the HITRAN2020 list. To clarify the trends, Fig.~\ref{fig:kfac} has a zoomed-in y-axis that excludes most outliers. The offset of about 180~m\,s$^{-1}$ between the spectra observed in April 2020 to the ones observed in July 2020 is due to a readjustment of the set-up. Such wavenumber offsets can reach ranges up to km\,s$^{-1}$. Another conspicuous pattern to note in Fig.~\ref{fig:kfac} is that the velocities of the single spectra are not stable over wavelengths as they are expected to be. The first panel shows the most stable velocities in the wavelength range between 6270 and 6350~\r{A}. Compared to the first panel, the second panel shows increased velocities for the wavelength range between 6850 and 7000~\r{A}. For all four observations the velocities raise about 100~m\,s$^{-1}$ but show no constant velocity overall. Nevertheless, the pattern of velocities is arranged in a similar and stable way for each observation. The last panel shows the correction velocities of the range between 7680 and 7750~\r{A}. The velocities in this segment are spread over a range of about 150~m\,s$^{-1}$. Like in the second panel, we see a similar pattern for all observations in this wavelength range. We found this consistent pattern for all observations, which indicates that these velocity leaps are a systematic pattern of the HITRAN2020 reference line list (all reference line positions originate from the model of \cite{2014JChPh.141q4302Y}) and not originating from inaccuracies of measurements or line determination. Since the wavelength region between 6270\,\r{A} and 6350\,\r{A} shows the most consistent velocities, we chose the telluric lines in this area for our wavenumber correction. To check for consistency, we used three deep O$_2$ lines in the wavelength range between 6300\,\r{A} and 6306\,\r{A} and we cross-checked our calibrated tellurics with observations of the Laser Absolute Reference Spectrograph (LARS), where an absolute reference comb was used for wavelength calibration. The difference between our corrected tellurics and the tellurics measured by the laser comb is at an average of $\approx$ 10~m\,s$^{-1}$. For the standard deviation of the LARS tellurics we determine about 20~m\,s$^{-1}$, while we receive 3.6~m\,s$^{-1}$ for the corrected IAG tellurics. Since the average difference between the two observations is within one standard deviation of the absolute reference tellurics we decide to use the HITRAN2020 lines between the wavelength range 6270 and 6350~\r{A} as the best option for our wavenumber correction. This wavelength range includes 45 telluric lines to calculate $\kappa$ from their median shift. We corrected every spectrum individually since $\kappa$ is unique for every FTS scan.

Another important fact about using tellurics as a reference is that they hold an intrinsic pressure shift. The pressure shifts provided by the HITRAN2020 database are usually on the order of about 50 to 100~m\,s$^{-1}$ and are wavelength-dependent. To calculate the accurate pressure shift for our lines, we would have to calculate a scaling factor for this velocities. However, since Earth's atmosphere is not stable and homogeneous calculation of this factor is not possible for our case \citep[see][]{2020ApJS..247...24B}, we applied the calibration without this pressure shift, as done by \cite{2016A&A...587A..65R}.

\subsubsection{Telluric contamination}
Ground-based spectral observations always have to deal with atmospheric line contamination by Earth's atmosphere. These telluric lines are usually stable in wavelength on the order of 10s of m/s \citep[see e.g.][]{2010A&A...515A.106F}. However, since we also have to correct our observed spectra for rotational velocities of the Sun as well as for the velocities between Earth and Sun, for the time of observation, the spectra get shifted respectively. Accordingly, the tellurics of the averaged spectra would broaden to hundreds of m/s. This broadening forces us to eliminate these lines in the first place. We use the HITRAN2020 database to get the specific wavelengths of H$_2$O and O$_2$ lines in Earth's atmosphere and exclude these areas from our spectra. 

\subsubsection{Relative velocity correction}

The next important calibration step is subtracting the relative motion between the VVT and the respective solar position from our spectra. We used an ephemeris code from \cite{2015PhDT.......200D}\footnote{\url{https://github.com/haped/solarv}}, which is based on NASA's Navigation and Ancillary Information Facility SPICE toolkit \citep{1996P&SS...44...65A} and has an accuracy of a few cm/s. The relative velocity of the motion between the Sun and Earth is subtracted from every observed spectrum by the appropriate spectral shift. The gravitational red-shift of the Sun related to the position of Earth (633~m\,s$^{-1}$) is not taken into account in this data calibration. To combine all spectra of the same radius $\mu$, it is also necessary to subtract the differential solar rotation. To determine the rotation velocity at the observed position we use the rotation law of the Sun described by a model of \cite{1990ApJ...351..309S}. The angular sidereal velocity $v$ in degrees per day is calculated as
\begin{align}
        v(l)= 14.714 - 2.396 \sin^2 l - 1.787 \sin^4 l,
\end{align}
\noindent where $l$ stands for the solar latitude. For the correction we use the rotation coefficients of the spectroscopic rate from \cite{1990ApJ...351..309S}.

\begin{figure}
        \centering
        \includegraphics[width=0.49\textwidth]{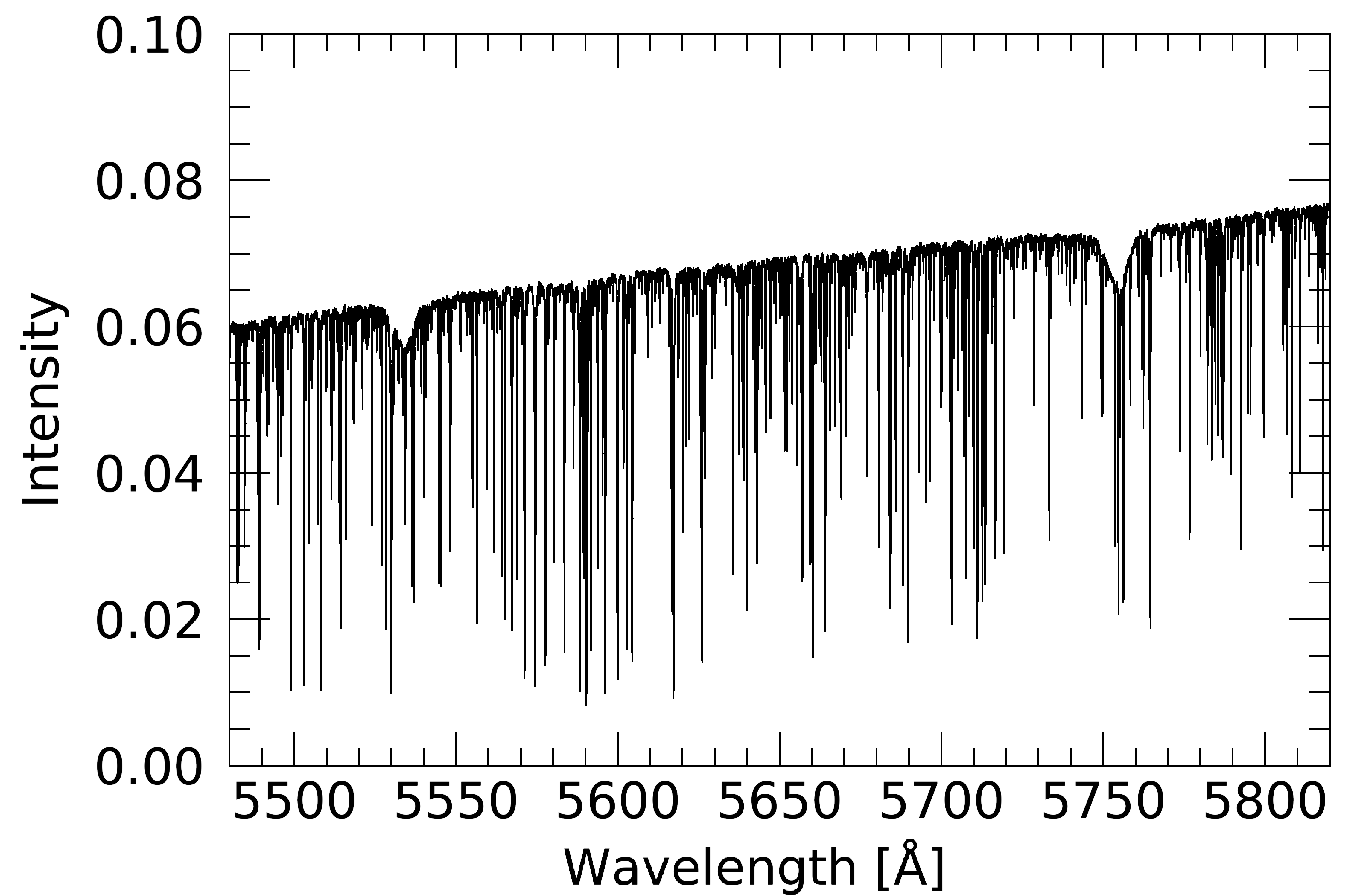}
        \caption{Solar spectra observed with the dichroic filter before normalisation. At 5535~\r{A} and 5755~\r{A} absorption features of the filter are visible.}
        \label{fig:dichroic_spect}
\end{figure}

\subsubsection{Normalisation} \label{norm}
The last step of our data processing is to normalise the continuum of the spectra to unity. For this purpose, an iteration loop is used to identify the upper local 10\% of intensity flux and fit a spline to the received continuum. Dividing the spectra by this continuum spline results in the normalised spectra. In the first half of our observation campaign (83 observations), we used a dichroic filter to remove the wavelengths over $\approx$ 8000~\r{A} from the FTS scan. We decided to remove the filter from the set-up after we observed notches in the solar spectra which originate from this filter. In Fig. \ref{fig:dichroic_spect} two of these notches are shown. Since single notches vary in wavelength for up to 30~\r{A} between different observation set-ups, we excluded 12 of these small areas from the data set to prevent false intensity variations.

\subsection{Error sources}

To get a sense of the quality of our data, we examined various error sources and accuracies of our data and list them in Table\,\ref{tab:values}. In addition, we investigated the effect of an extended observation area on the solar surface due to the fibre, instead of a point source.

\subsubsection{Signal-to-noise ratio}
To compare the level of the spectral signal to the level of our background noise, we calculated the signal-to-noise ratio (S/N). We estimated the spectral noise $\sigma_{\rm n}$ for a single observation $F_{n, \nu}$ by comparison with the averaged spectrum $F_{\nu}$ 
\begin{align}
    \sigma_n = {\rm rms}\left(F_{n,\nu} - F_{\nu}\right).
\end{align}
The spectral range we use for this calculation is about 2.2\,{\AA} in the continuum ($\lambda = 6676.4 - 6678.6$\,\AA). The S/N is computed from $\sigma_n$, with
\begin{align}
    {\rm S/N} = \frac{F_n}{\sigma_n},
\end{align}
where $F_n = \langle F_{n,\nu} \rangle \approx 1$ is the mean of the corresponding spectral range.
Eventually, the values of the S/N depend on the observation radii $\mu$. We receive values from 420 ($\mu$ = 0.2) to over 600 at and close to the disc centre for single spectra and per pixel (see Table~\ref{tab:values}). A pixel is defined as $\Delta \nu = 0.0075~\text{cm}^{-1}$, which is a third of the resolution element.

\subsubsection{Pointing accuracy}

A decent pointing accuracy is needed to determine the spectrum's variation over the solar disc. The uncertainty of our pointing on the solar surface depends on the selected position on the Sun. From the centre of the disc, with a pointing error $\delta(r)$ of about 5\arcsec~, it rises to $\delta(r)$ of about 12\arcsec~on the solar limb. For a more detailed description about these uncertainties and their origin, we refer to \cite{10.1117/12.2560156}. The most affected position at $\mu$ = 0.2 includes a maximal pointing error range between $\mu$ = 0.16 - 0.23. Observing a spectrum at another $\mu$ position than wanted leads to a different spectrum in general and will also be wrongly corrected in terms of relative velocities since the rotational velocity will be inaccurate. The rotational velocity of the $\mu$ radii 0.16 to 0.23 differs within $\Delta v_{point.}$ = 22~m\,s$^{-1}$, where the maximal velocity difference corresponds to the rotational axis. The effect of the maximal pointing error of the east-west axis of the Sun is listed in Table~\ref{tab:values} for all radii. These values represent the maximal velocity offset, since the solar rotation has the strongest impact on these kind of errors. For other disc positions than those along the rotational axis, this offset gets smaller.

\subsubsection{Illumination of the fibre} \label{fibre}

Since our fibre observes an area of 32.5" in diameter on the solar surface, we always observe a superposition of different $\mu$ radii. To check which amount of the observed light, especially at the solar limb, truly originates from the desired observing position, we need to understand the impact of this superposition in more detail. Therefore, we simulated the illumination of the fibre for our solar observations. For the sake of simplicity, we calculated the fibre as a circle instead of the actual hexagonal shape. The difference in surface area is about 17\% between the two shapes, but there is no possibility to check the orientation of the fibre inside the set-up.
\begin{figure}
        \centering
        \includegraphics[width=0.45\textwidth]{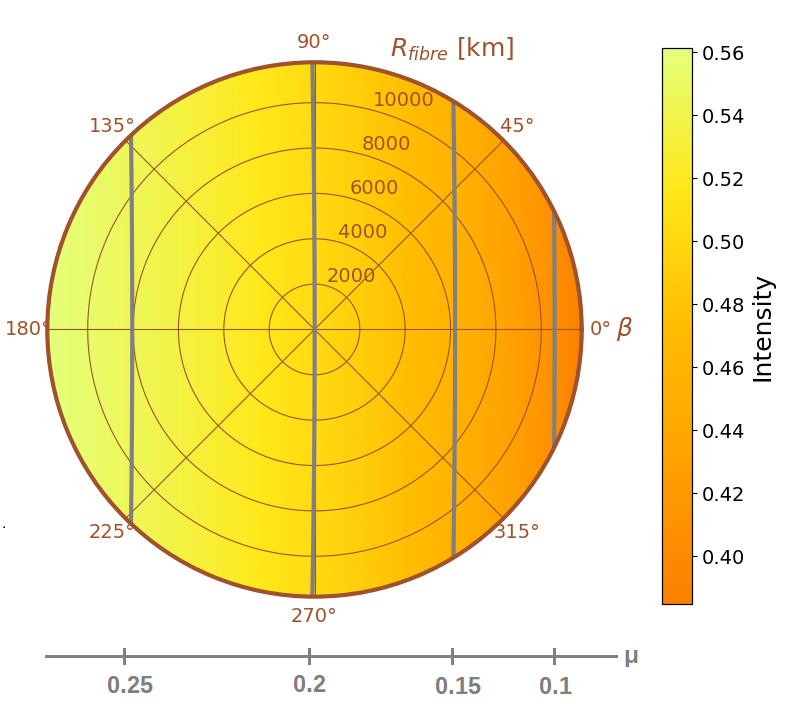}
        \caption{Fibre illumination at pointing position $\mu_{\rm pos} =$ 0.2. The centre of the Sun is directed towards $\beta$ = 180$^\circ$. The grey lines mark the $\mu$ radii of 0.25, 0.2, 0.15, and 0.1 from left to right.}
        \label{fig:fibre}
\end{figure}
\begin{figure}
        \includegraphics[width=0.4\textwidth]{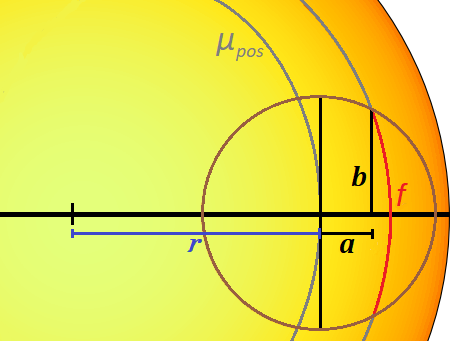}
        \caption{Sketch (not to scale) to support Eq.~\ref{eq:flux_dis}. With $r$ as the distance between the solar centre and the centre of the fibre. The red arc length $f$ marks the contribution of a specific $\mu$ position of one observation.}
        \label{fig:sketch}
\end{figure}
\begin{table}
        \centering
        \small
        \caption{$\mu_{\rm pos}$ with characteristic values.}
        \begin{tabular}{@{}rrrr@{}}
                \hline
                \hline
                $\mu_{\rm pos}$ & $\mu_{\rm wm}$ & $\mu_{\rm min}$ & $\mu_{\rm max}$\\
                \hline
                1.00 & 0.99995 & 0.999 & 1.000 \\
                0.95 & 0.94997 & 0.944 & 0.955 \\
                0.90 & 0.89998 & 0.892 & 0.908 \\
                0.80 & 0.80002 & 0.787 & 0.812 \\
                0.70 & 0.70009 & 0.682 & 0.717 \\
                0.60 & 0.60022 & 0.577 & 0.622 \\
                0.50 & 0.50049 & 0.469 & 0.528 \\
                0.40 & 0.40110 & 0.359 & 0.437 \\
                0.35 & 0.35177 & 0.301 & 0.392\\
                0.30 & 0.30294 & 0.240 & 0.349\\
                0.20 & 0.21041 & 0.081 & 0.270\\
                \hline
        \end{tabular}
    \tablefoot{$\mu_{\rm pos}$ positions, the calculated weighted mean $\mu_{\rm wm}$ and their corresponding $\mu_{\rm min}$ and $\mu_{\rm max}$ values to Fig.\,\ref{fig:fibre_over_mu}.}
\label{tab:mus}
\end{table}             

Figure~\ref{fig:fibre} illustrates the illumination of the fibre for the pointing position at $\mu$ = 0.2. This radius is the most affected $\mu$ value, with superpositions of radii between 0.08 and 0.27. To calculate the respective solar flux inside the fibre, we used the intensity profile of the given limb-darkening function from \cite{1994SoPh..153...91N} at the wavelength $\lambda$ = 5800\,\r{A} and normalised it to the intensity of the centre of the solar disc. The amount of the arc length at certain limb radii $\mu$, if the fibre is pointed at $\mu_{\text{pos}}$, is given by
\begin{align}
        \label{eq:flux_dis}
   f(\mu, \mu_{\rm pos}) = 2\cdot \mu \cdot \text{arctan}\Biggl(\frac{b(\mu)}{r(\mu_{\rm pos})+a(\mu)} \Biggr),
\end{align}
where $\mu_{\text{pos}}$ is the pointing position of the centre of the fibre and  $\mu$ are the respective radii for this pointing position. Figure~\ref{fig:sketch} shows a sketch of the fibre on the solar surface where the variables of Eq. \ref{eq:flux_dis} are explained.

\begin{figure}
        \centering
        \includegraphics[width=0.49\textwidth]{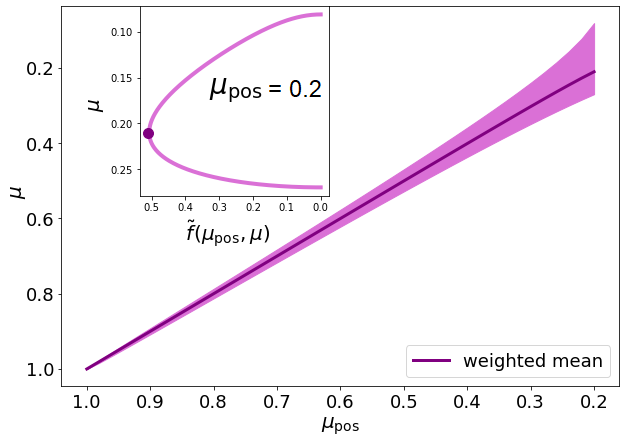}
        \caption{Contribution of $\mu$ radii contained at respective $\mu_{\rm pos}$ observation radii. The broad band shows the total distribution of $\mu$, the darker line the weighted mean of a specific radius. The cross-section inset shows the weighted flux distribution $\widetilde{f}(\mu, \mu_{\rm pos})$ for $\mu_{\rm pos}$ = 0.2, where the weighted mean is marked by a dot.}
        \label{fig:fibre_over_mu}
\end{figure}

The contribution of different $\mu$ radii for observations at $\mu_{\text{pos}}$ can be seen in Fig.~\ref{fig:fibre_over_mu}. The effect of the mixing $\mu$ radii gets stronger for smaller $\mu_{\rm pos}$ values since the $\mu$ radii do not change linearly.
For the case of $\mu_{\rm pos} = 0.2,$ we show the weighted flux distribution, $\widetilde{f,}$ over $\mu$ in the smaller plot inside Fig.~\ref{fig:fibre_over_mu}. This weighted flux $\widetilde{f}$, which is originating from the arc length $f$, is involving the intensity changes of the centre-to-limb effect. There, one can see the weighted intensity distribution in terms of the disc centre with slightly higher intensities for larger $\mu$ values, since the disc is brighter in its centre. The visible asymmetry originates from observing a sphere with non-equivalent $\mu$ steps and because of the presence of limb darkening.
This distribution shows the effect of the fibre size itself. The spectra towards the limb of the solar disc become more and more a superposition of different $\mu$ radii. Table~\ref{tab:mus} lists the positions of the weighted means $\mu_{\rm wm}$ of the weighted intensity distribution, $\widetilde{f,}$ for $\mu_{\rm pos}$ and also the extreme minimum and maximum $\mu$ values included in the fibre observations. The difference between the desired position $\mu_{\rm pos}$ and the weighted mean $\mu_{\rm wm}$ of the distribution is close to zero for most radii. For $\mu_{\rm pos} = 0.2,$ the fibre range includes $\mu \approx (0.08 - 0.27)$, which leads to the most displaced weighted mean value of $\mu_{\rm wm} = 0.21$.

As discussed in this sub-section, our set-up is limited to the uncertainty of the fibre size itself. Even if the fibre illumination at close limb positions is irregular, we find minor differences between the weighted mean and $\mu_{\rm pos}$. Nevertheless, the spectra we observe are still a superposition of all the radii inside the fibre. This superposition leads to a general spectral line smearing, especially towards the solar limb. 

\subsubsection{Error budget}
In the previous sub-sections, we explain the data processing and error sources of our data set. We also give a conclusive overview of the data quality. In Table~\ref{tab:values}, we list the quality characteristics for each $\mu$ radii. To calculate the wavelength precision, we determined the offset between the averaged spectra and every single spectrum per $\mu$. We used the wavelength range between 6100 and 6180~\r{A} to calculate the offset radial velocity shift between the spectra. The results of this calculations are listed as $\sigma_{\rm scat}$ in Table~\ref{tab:values}.

Additionally, a velocity offset $\Delta v_{\rm rot.}$ is listed, which is the maximal rotational velocity offset between the pointing position of our fibre, $\mu_{\rm pos}$, and the actual weighted mean position, $\mu_{\rm wm}$, described in Sect.\,\ref{fibre}. This maximum is calculated on the east-west axis of the Sun, where the solar rotation is strongest. As a result, the $\Delta v$ value for the averaged spectra is always smaller than $\Delta v_{\rm rot.}$. For the disc centre at $\mu_{\rm pos} = 1 $, this effect is neglectable, since the $\mu$ values are arranged in circles around the centre and the contrary rotational directions cancel each other out. For $\mu=0.2,$ with the broadest divergence to $\mu_{\rm wm} = 0.21,$ the calculated rotational offset is 4.02\,m\,s$^{-1}$.

To estimate the accuracy of our data, we used the \ion{Fe}{I} $\lambda_0 = 6175$\,{\AA} (vacuum wavelength) solar line. \cite{2019A&A...622A..34S} measured this line for different $\mu$ positions, using a laser comb obtaining an uncertainty of m\,s$^{-1}$. Assuming the line positions of \cite{2019A&A...622A..34S} as perfect, the deviation between their line position and the line position of the IAG atlas leads to an estimate for our line accuracy of about 16\,m\,s$^{-1}$ for $\mu = 1$. The line accuracy depends on the $\mu$ radius. In Table~\ref{tab:values} the deviations between both observed data sets are listed as $\Delta v_{{\rm LARS},6175}$. The list is empty at $\mu$ positions where \cite{2019A&A...622A..34S} does not provide any data. All values are below about 31\,m\,s$^{-1}$, except for the position of $\mu=0.95$  where we measure a divergence of 48.56\,m\,s$^{-1}$. We notice higher accuracies (smaller velocity) for IAG $\mu$ radii with a higher total observation time.

\section{Results} \label{Results}

In this section, we present selected results on the spectral centre-to-limb behaviour of our solar observations. We picked the prominent \ion{Fe}{I} 6175~\r{A} (vacuum wavelength) absorption line and compared the IAG data with observations of LARS \citep[see][]{2017A&A...607A..12L,2019A&A...622A..34S,2019A&A...624A..57L} to check the consistency of these two benchmark observations. Even if the \ion{Fe}{I} 6175~\r{A} is magnetically sensitive, since we are observing the quiet Sun regions and adding up a range of observations, we assume that the impact of accidentally observed magnetic areas is negligible. Additionally, we counter-checked our results with the 3D radiative MHD code MURaM \citep[see][]{2005A&A...429..335V,2013A&A...558A..49B,2015A&A...581A..43B}. We used the non-magnetic G2V-hydro model (cf. Table~1. in \cite{2015A&A...581A..43B}) of a grid of 512 × 512 horizontal and 800 vertical cells, which corresponds to a 9-9-3~Mm box size. For the spectral line synthesis, the forward module of the Spinor code \citep{2000A&A...358.1109F} was used.

The LARS data are from an observation campaign in 2016 at the Vacuum Tower Telescope in Tenerife, with a spectral resolution of $\lambda / \Delta \lambda$ > $700,000$ at 6175~\r{A}. The pointing accuracy of the telescope is given with about 1". Their spectra contain an S/N from 800 to 1000 at the disc centre. The LARS data we use for this work represent the average of all their measurements per $\mu$ radius. Each observation sequence is about 20~min, leading to a total observing time of 180~min for the disc centre and 60~min or 80~min for the observations towards the solar limb. Unlike our IAG observations, the LARS observations differ in size of the observed disc areas for changing $\mu$ radii. To mitigate the increasing effect of supergranulation towards the solar limb the observed area was varied from 40"$\times$10" (inner disc) to 40"$\times$30" (solar limb). The limb observation regions are similar in size to the fibre size of the IAG set-up. Also, the area for $\mu$ = 1 is on the same order for both observations. More details about the observations and the data can be found in \cite{2018A&A...611A...4L}. 

\begin{figure}
        \centering
        \includegraphics[width=0.49\textwidth]{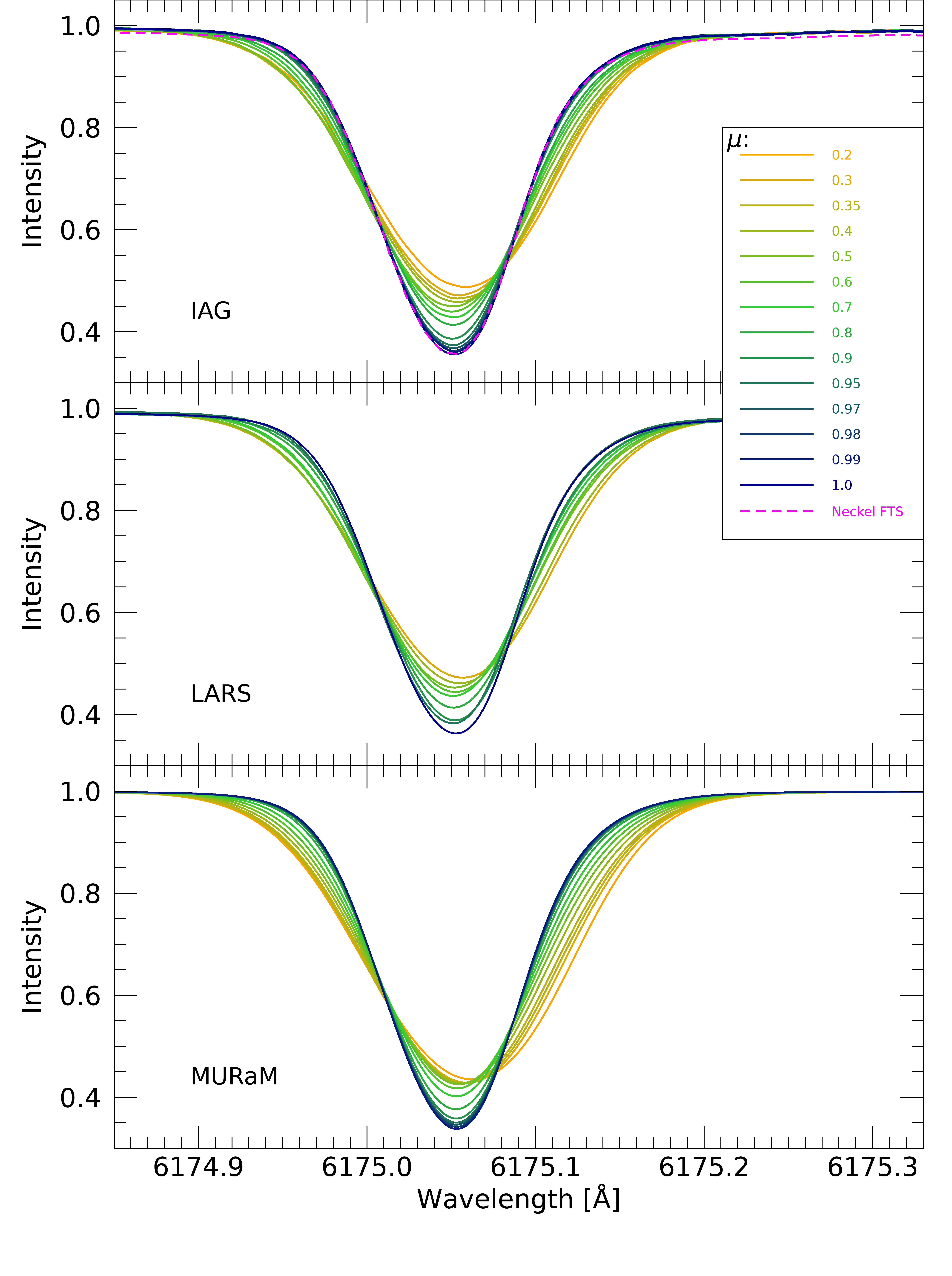}
        \caption{Line profiles of different $\mu$ radii for the Fe I 6175\,\r{A} line in vacuum wavelength. Upper panel: IAG data for the average of the respective $\mu$ angle plus to the Neckel FTS atlas for $\mu$=1.0. Middle panel: LARS observations. Lower panel: MURaM MHD simulations.}
        \label{fig:iag_mu}
\end{figure}

In Fig.~\ref{fig:iag_mu}, the profiles for different $\mu$'s of the FeI 6175~\r{A} line are shown for IAG, LARS, and the simulated MURaM data. For comparison, we show the disc-centre Neckel FTS atlas \citep{1999SoPh..184..421N} for this absorption line. The Neckel line shows an excellent agreement with the observations at $\mu$ = 1.0. We found no systematic divergence and we computed a difference of 0.9\% of the total intensities between the IAG and Neckel line.

The first point to notice is a steadily decreasing line depth for decreasing $\mu$ values as a general behaviour for all cases.
For the MURaM lines, this behaviour seems to be less pronounced for $\mu$ values smaller than $\approx$~0.5, while IAG and LARS data show a clear ongoing process in line depth change for all radii. Additionally, the spectral lines get broader and shifted to higher wavelengths for decreasing $\mu$ for all three data sets.

\begin{figure}
        \centering
        \includegraphics[width=0.49\textwidth]{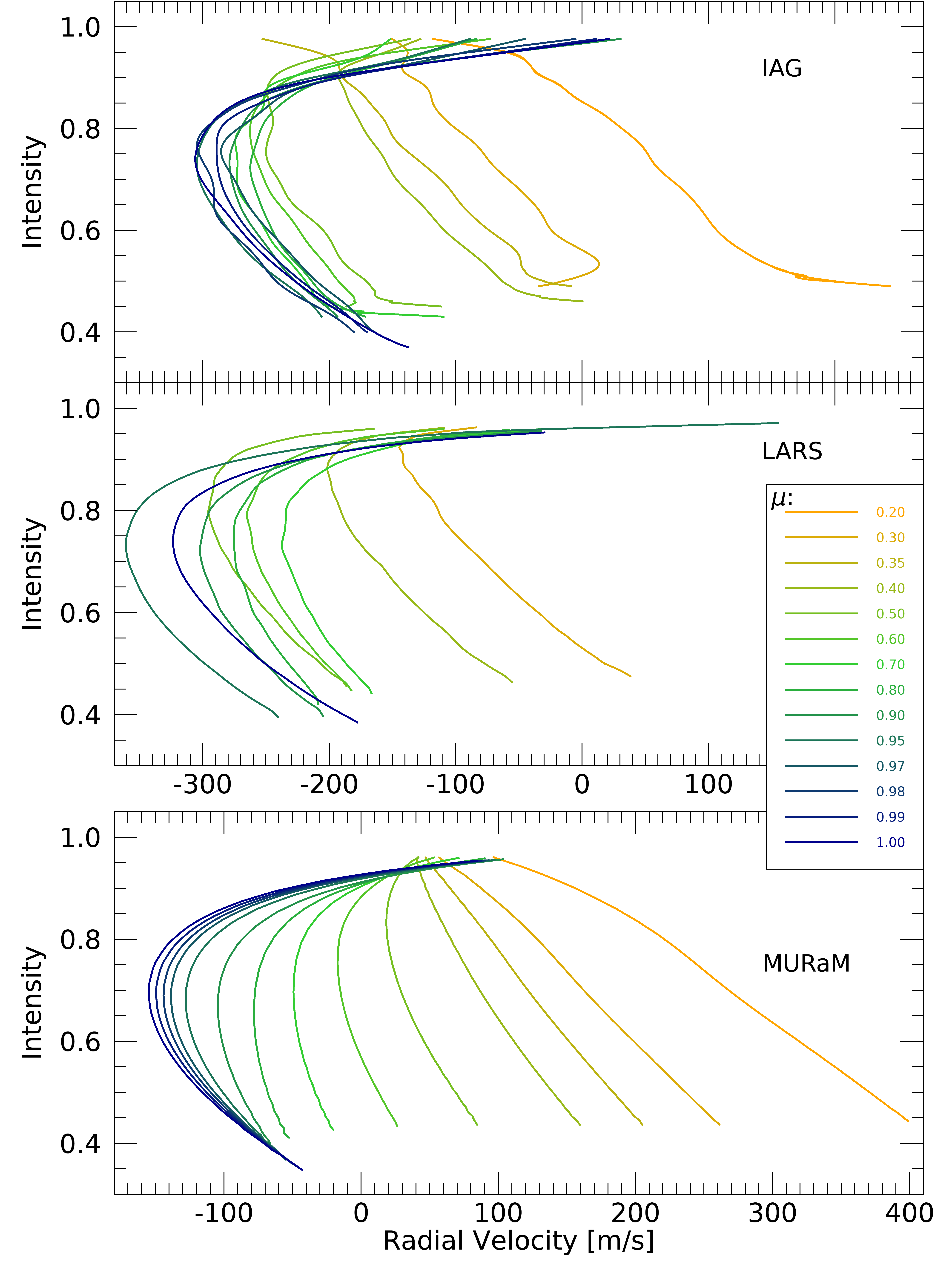}
        \caption{Bisectors of different $\mu$ radii for the Fe I 6175~\r{A} line. Upper panel:\ IAG data. Middle: LARS data for comparison. Lowest panel:\ Bisectors from the MURaM simulated data.}
        \label{fig:spec}
\end{figure}

To observe subtle differences in line shapes in more detail, we determine the bisectors of the lines in terms of radial velocity. For the reference wavelength we use $\lambda_{\rm vac} = 6175.0428\,$~\r{A}, as they used the same wavelength in \cite{2019A&A...622A..34S}. In Fig. \ref{fig:spec} the bisectors corresponding to Fig.~\ref{fig:iag_mu} are shown. The bisectors of the IAG and LARS seem to be in good accordance with velocity behaviour and shape for most $\mu$ values. The bisectors for smaller $\mu$ radii follow the same trend while the IAG bisectors of higher $\mu$ values are more clustered than the LARS data. The bisectors strongly shift to higher velocities in both data sets for solar limb positions of $\mu$ = 0.4 and smaller. The LARS observations did not include the radii of $\mu$ = 0.2, 0.35, 0.97, 0.98, and 0.99.

The main differences between the two observations can be seen at $\mu$ = 0.95, 0.7, and 0.5. For $\mu$ = 0.95, the LARS observations show the strongest blue-shift with about $-$360\,m\,s$^{-1}$ at an intensity level of about 0.75. At the disc centre ($\mu$ = 1), the LARS data only show a blue-shift of $-$320\,m\,s$^{-1}$ at the same intensity level. The IAG data shows the strongest blue-shift also at an intensity level of $\approx$ 0.75 but with about $-$305\,m\,s$^{-1}$ for all the radii of 1.0, 0.98 and 0.95. For $\mu$ = 0.95 and 0.8, the LARS observations were still scanning  a surface area of 40\arcsec$\times$10\arcsec, while the IAG set-up always observes 32.5" in diameter, which could be a reason for the mismatch between these RVs. In addition, the IAG spectra for $\mu$ = 0.5 and 0.7 have a relatively short observing time compared to the other observations. With just 80~min (resp. 50~min) total observing time this could, for example, lead to spectra that are still biased due to supergranulation.

The last panel of Fig.~\ref{fig:spec} shows the bisectors of the simulation. Compared to the observations they are shifted by about 150~m\,s$^{-1}$ to higher velocities for $\mu$ = 1 because the whole wavelength range of the line is already shifted. The greatest blue-shift is visible for $\mu$ = 1 with the strongest blue-shift of -158\,m\,s$^{-1}$ at an intensity level of 0.7. The radial velocity continuously increases for decreasing $\mu$. In particular, the bisectors for solar radii of $\mu$ = 0.5 up to 0.95 show a different line shape than the observed bisectors. \cite{2019A&A...624A..57L} investigated the centre-to-limb behaviour of 26 absorption lines observed by LARS. They also compared blue-shift variations over the solar disc with radiative transfer LTE computations of non-magnetic 3D hydrodynamical photospheric simulations.  Overall, this examination results in deviations between simulations and observations by less than 100\,m\,s$^{-1}$. However, the simulations often could not reproduce the correct behaviour of the centre-to-limb variation. In comparison, the calculations of this work result in velocity deviations between observations and simulations of up to 200\,m\,s$^{-1}$.

\cite{2015ApJ...808...59K} used 3D radiative hydrodynamic simulations to investigate characteristics of the \ion{Fe}{I} 6175~\r{A} line for four observing angles ($\mu$ = [1, 0.87, 0.71, 0.5]). The authors found a Doppler shift of about -50 to -100\,m\,s$^{-1}$ for the disc centre and a red-shift of about 100 to 210\,m\,s$^{-1}$ for $\mu$ = 0.5. These values are more similar to MURaM than to the observations of IAG and LARS. The line cores of MURaM show velocities of -40\,m\,s$^{-1}$ for the disc centre and 90\,m\,s$^{-1}$ for $\mu$ = 0.5. However, the total trend seems to differ, since the velocities of MURaM are negative from $\mu$ = 0.7 up to 1.0 while the simulations of \cite{2015ApJ...808...59K} show already positive velocities close to the disc centre at $\mu$ = 0.95.

\begin{figure}
        \centering
        \includegraphics[width=0.49\textwidth]{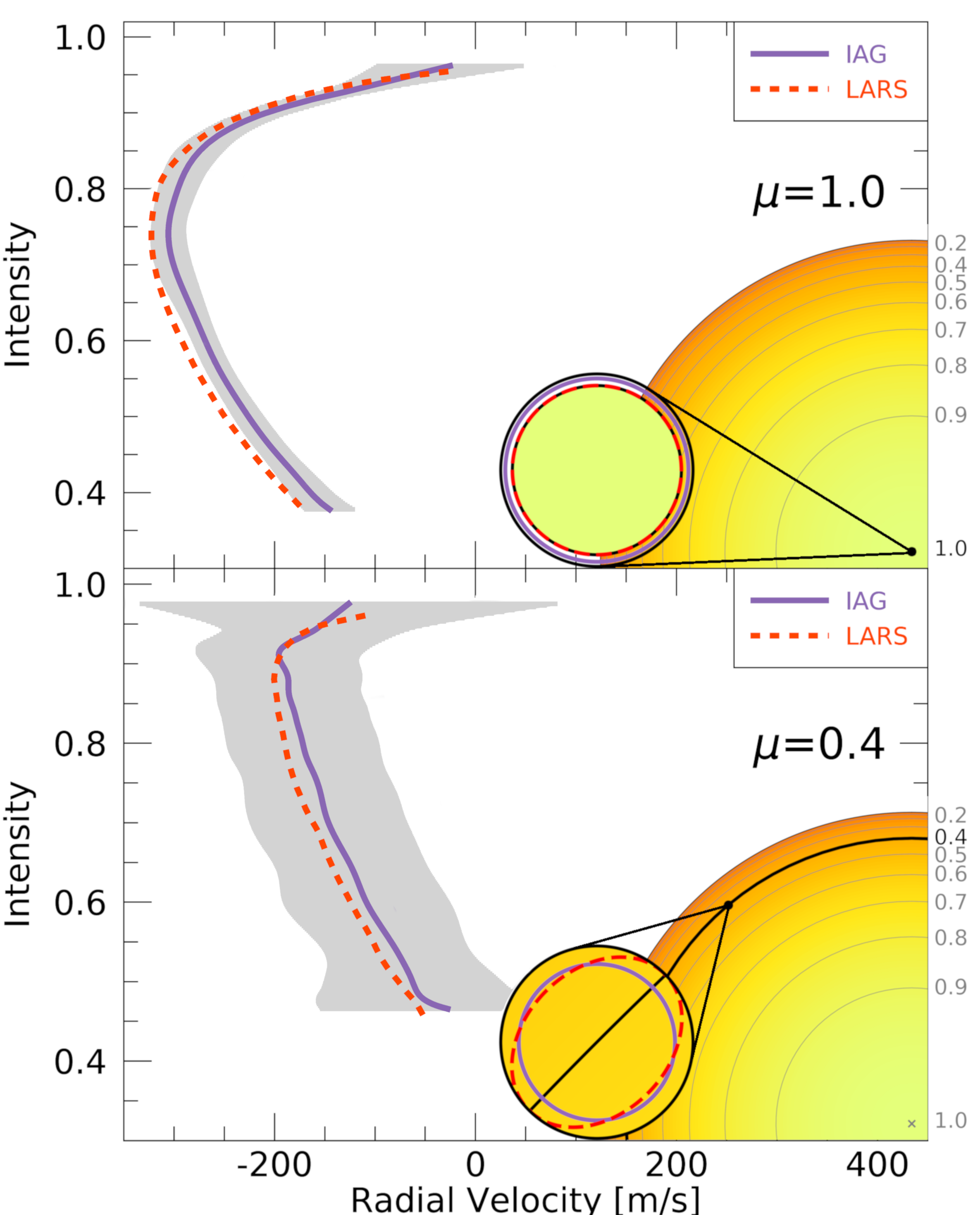}
        \caption{Bisectors of $\mu$=1.0 and $\mu$=0.4 for the 6175~\r{A} line with the corresponding observation area. The grey area marks the standard deviation of the IAG data. The purple circle marks the size of the IAG fibre with 32.5", while the red areas mark the LARS observations.}
        \label{fig:mu10}
\end{figure}

For a better comparison, we show the bisectors of the IAG and LARS data for the radii of $\mu$ = 1.0 and 0.4 in Fig.~\ref{fig:mu10} together in one plot. Also, the standard deviation of the averaged IAG spectra is shown. This plot makes it clearer how accurately the IAG and LARS bisectors match. It also shows a small offset of about 15~m\,s$^{-1}$. All $\mu$ radii plots can be found in Appendix \ref{ap:fig}. We recognise a blue-shift for the IAG bisectors compared to the LARS data for every $\mu$ value, but $\mu = 0.7$. This shift is on the order between 15~m\,s$^{-1}$ for the disc centre and a maximum of about 60~m\,s$^{-1}$ for $\mu = 0.95$. As mentioned earlier in this paper, $\mu$ = 0.7 has a shorter observation time (50~min) than other radii, which could be the reason for an intrinsic offset.  

\section{Summary} \label{Summary}

In this article, we introduce the centre-to-limb IAG solar atlas, including the wavelength range from 4200~\r{A} to 8000~\r{A} for 14 heliocentric positions of the solar disc. An electronic version of the atlas is available. We visualised our data online\footnote{\url{ https://www.astro.physik.uni-goettingen.de/research/solar-lib}}, including the entire spectra. We examine the data in more detail in follow-up analysis that is beyond the scope  of this paper. Here, we provide an overview of the instrumental and observational characteristics of our data and give our preliminary insights.
We investigated the centre-to-limb behaviour of the \ion{Fe}{I} 6175~\r{A} line and compared our atlas to previous centre-to-limb observations of the LARS instrument as well as to simulated lines using 3D MHD simulations. We generally ascertain a similar behaviour for the two observed data sets with a velocity shift between approximately 15-60\,m\,s$^{-1}$. The IAG data shows consistency with the LARS observations for most $\mu$ radii with a blue-shift in the range of 15-30\,m\,s$^{-1}$. For $\mu$ = 0.95, we found the largest velocity offset between IAG and LARS with 60\,m\,s$^{-1}$. At this disc position, LARS observations were carried out for a smaller surface area than the IAG observations. This velocity offset could reveal a sensitive atmospheric structure for these observing radii, where the spectrum is quite sensitive to the superposition of different $\mu$ radii. Furthermore, the LARS dataset originates from 2016 observations, which were closer to a solar maximum than the IAG observations. \cite{2010A&A...512A..39M} and \cite{2016A&A...587A.103L} described a velocity variation of about 5 to 10\,m\,s$^{-1}$ over the solar cycle on the integrated Sun, which could also lead to an intrinsic offset between the two datasets.

The bisectors of the simulations vary distinctly in velocity and shape from observations, especially towards the solar limb. Also, the comparison between results of different simulations leads to a high variance in the line behaviour. This difference implies that there is potential to use this kind of data to improve atmospheric simulations.

This high-resolution spatially resolved solar spectral atlas includes a spectral range of thousands of\,\r{A} in the visible. As a result, it will improve our knowledge of the Sun's atmospheric structure and its spectral centre-to-limb fingerprint. This novel atlas serves as benchmark for a broad field of solar and stellar physics and will be useful for detailed solar limb investigations in the future, understanding solar RV jitter in general, and further improvements to 3D MHD simulations.

\begin{acknowledgements}
    We thank the anonymous referee for valuable comments and suggestions which helped to improve this manuscript. We would like to thank J. L\"ohner-B\"ottcher and W. Schmidt to provide us with the LARS data and for fruitful discussions, H-P Doerr for helping to install and use the ephemeris code and the lab-team at IAG for technical support.   ME acknowledges financial support through the SPP1992 under Project DFG RE 1664/17-1. The FTS was funded by the DFG and the State of Lower Saxony through the Gro{\ss}ger\"ateprogramm Fourier Transform Spectrograph. 
\end{acknowledgements}

\bibliographystyle{aa}
\bibliography{SolarSpectrumFTS}

\begin{thebibliography}{49}
\expandafter\ifx\csname natexlab\endcsname\relax\def\natexlab#1{#1}\fi

\bibitem[{{Acton}(1996)}]{1996P&SS...44...65A}
{Acton}, C.~H. 1996, \planss, 44, 65

\bibitem[{{Allende Prieto} {et~al.}(2004){Allende Prieto}, {Asplund}, \&
  {Fabiani Bendicho}}]{2004A&A...423.1109A}
{Allende Prieto}, C., {Asplund}, M., \& {Fabiani Bendicho}, P. 2004, \aap, 423,
  1109

\bibitem[{{Baker} {et~al.}(2020){Baker}, {Blake}, \&
  {Reiners}}]{2020ApJS..247...24B}
{Baker}, A.~D., {Blake}, C.~H., \& {Reiners}, A. 2020, \apjs, 247, 24

\bibitem[{{Balthasar}(1988)}]{1988A&AS...72..473B}
{Balthasar}, H. 1988, \aaps, 72, 473

\bibitem[{{Beckers} \& {Nelson}(1978)}]{1978SoPh...58..243B}
{Beckers}, J.~M. \& {Nelson}, G.~D. 1978, \solphys, 58, 243

\bibitem[{{Beeck} {et~al.}(2013){Beeck}, {Cameron}, {Reiners}, \&
  {Sch{\"u}ssler}}]{2013A&A...558A..49B}
{Beeck}, B., {Cameron}, R.~H., {Reiners}, A., \& {Sch{\"u}ssler}, M. 2013,
  \aap, 558, A49

\bibitem[{{Beeck} {et~al.}(2015){Beeck}, {Sch{\"u}ssler}, {Cameron}, \&
  {Reiners}}]{2015A&A...581A..43B}
{Beeck}, B., {Sch{\"u}ssler}, M., {Cameron}, R.~H., \& {Reiners}, A. 2015,
  \aap, 581, A43

\bibitem[{{Bergemann} {et~al.}(2021){Bergemann}, {Hoppe}, {Semenova},
  {Carlsson}, {Yakovleva}, {Voronov}, {Bautista}, {Nemer}, {Belyaev},
  {Leenaarts}, {Mashonkina}, {Reiners}, \& {Ellwarth}}]{2021MNRAS.tmp.1964B}
{Bergemann}, M., {Hoppe}, R., {Semenova}, E., {et~al.} 2021, \mnras

\bibitem[{{Bray} \& {Loughhead}(1967)}]{1967sogr.book.....B}
{Bray}, R.~J. \& {Loughhead}, R.~E. 1967, {The solar granulation}

\bibitem[{{Crass} {et~al.}(2021){Crass}, {Gaudi}, {Leifer}, {Beichman},
  {Bender}, {Blackwood}, {Burt}, {Callas}, {Cegla}, {Diddams}, {Dumusque},
  {Eastman}, {Ford}, {Fulton}, {Gibson}, {Halverson}, {Haywood}, {Hearty},
  {Howard}, {Latham}, {L{\"o}hner-B{\"o}ttcher}, {Mamajek}, {Mortier},
  {Newman}, {Plavchan}, {Quirrenbach}, {Reiners}, {Robertson}, {Roy}, {Schwab},
  {Seifahrt}, {Szentgyorgyi}, {Terrien}, {Teske}, {Thompson}, \&
  {Vasisht}}]{2021arXiv210714291C}
{Crass}, J., {Gaudi}, B.~S., {Leifer}, S., {et~al.} 2021, arXiv e-prints,
  arXiv:2107.14291

\bibitem[{{Cubas Armas} \& {Fabbian}(2021)}]{2021ApJ...923..207C}
{Cubas Armas}, M. \& {Fabbian}, D. 2021, \apj, 923, 207

\bibitem[{{Davis} {et~al.}(2001){Davis}, {Abrams}, \&
  {Brault}}]{2001ftsp.book.....D}
{Davis}, S.~P., {Abrams}, M.~C., \& {Brault}, J.~W. 2001, {Fourier transform
  spectrometry}

\bibitem[{{Doerr}(2015)}]{2015PhDT.......200D}
{Doerr}, H.~P. 2015, PhD thesis, University of Freiburg

\bibitem[{{Dravins} {et~al.}(1981){Dravins}, {Lindegren}, \&
  {Nordlund}}]{1981A&A....96..345D}
{Dravins}, D., {Lindegren}, L., \& {Nordlund}, A. 1981, \aap, 96, 345

\bibitem[{{Dravins} {et~al.}(2017){Dravins}, {Ludwig}, {Dahl{\'e}n}, \&
  {Pazira}}]{2017A&A...605A..90D}
{Dravins}, D., {Ludwig}, H.-G., {Dahl{\'e}n}, E., \& {Pazira}, H. 2017, \aap,
  605, A90

\bibitem[{{Dravins} {et~al.}(2021){Dravins}, {Ludwig}, \&
  {Freytag}}]{2021A&A...649A..17D}
{Dravins}, D., {Ludwig}, H.-G., \& {Freytag}, B. 2021, \aap, 649, A17

\bibitem[{{Figueira} {et~al.}(2010){Figueira}, {Pepe}, {Lovis}, \&
  {Mayor}}]{2010A&A...515A.106F}
{Figueira}, P., {Pepe}, F., {Lovis}, C., \& {Mayor}, M. 2010, \aap, 515, A106

\bibitem[{{Frutiger} {et~al.}(2000){Frutiger}, {Solanki}, {Fligge}, \&
  {Bruls}}]{2000A&A...358.1109F}
{Frutiger}, C., {Solanki}, S.~K., {Fligge}, M., \& {Bruls}, J.~H.~M.~J. 2000,
  \aap, 358, 1109

\bibitem[{Gordon {et~al.}(2022)Gordon, Rothman, Hargreaves, Hashemi, Karlovets,
  Skinner, Conway, Hill, Kochanov, Tan, Wcisło, Finenko, Nelson, Bernath,
  Birk, Boudon, Campargue, Chance, Coustenis, Drouin, Flaud, Gamache, Hodges,
  Jacquemart, Mlawer, Nikitin, Perevalov, Rotger, Tennyson, Toon, Tran,
  Tyuterev, Adkins, Baker, Barbe, Canè, Császár, Dudaryonok, Egorov,
  Fleisher, Fleurbaey, Foltynowicz, Furtenbacher, Harrison, Hartmann, Horneman,
  Huang, Karman, Karns, Kassi, Kleiner, Kofman, Kwabia–Tchana, Lavrentieva,
  Lee, Long, Lukashevskaya, Lyulin, Makhnev, Matt, Massie, Melosso,
  Mikhailenko, Mondelain, Müller, Naumenko, Perrin, Polyansky, Raddaoui,
  Raston, Reed, Rey, Richard, Tóbiás, Sadiek, Schwenke, Starikova, Sung,
  Tamassia, Tashkun, {Vander Auwera}, Vasilenko, Vigasin, Villanueva, Vispoel,
  Wagner, Yachmenev, \& Yurchenko}]{GORDON2022107949}
Gordon, I., Rothman, L., Hargreaves, R., {et~al.} 2022, Journal of Quantitative
  Spectroscopy and Radiative Transfer, 277, 107949

\bibitem[{{Griffiths} \& {de Haseth}(2007)}]{2007ftis.book.....G}
{Griffiths}, P.~R. \& {de Haseth}, J.~A. 2007, {Fourier Transform Infrared
  Spectrometry, Second Edition}

\bibitem[{{Halm}(1907)}]{1907AN....173..273H}
{Halm}, J. 1907, Astronomische Nachrichten, 173, 273

\bibitem[{{Kitiashvili} {et~al.}(2015){Kitiashvili}, {Couvidat}, \&
  {Lagg}}]{2015ApJ...808...59K}
{Kitiashvili}, I.~N., {Couvidat}, S., \& {Lagg}, A. 2015, \apj, 808, 59

\bibitem[{{Lanza} {et~al.}(2016){Lanza}, {Molaro}, {Monaco}, \&
  {Haywood}}]{2016A&A...587A.103L}
{Lanza}, A.~F., {Molaro}, P., {Monaco}, L., \& {Haywood}, R.~D. 2016, \aap,
  587, A103

\bibitem[{{L{\"o}hner-B{\"o}ttcher} {et~al.}(2017){L{\"o}hner-B{\"o}ttcher},
  {Schmidt}, {Doerr}, {Kentischer}, {Steinmetz}, {Probst}, \&
  {Holzwarth}}]{2017A&A...607A..12L}
{L{\"o}hner-B{\"o}ttcher}, J., {Schmidt}, W., {Doerr}, H.~P., {et~al.} 2017,
  \aap, 607, A12

\bibitem[{{L{\"o}hner-B{\"o}ttcher} {et~al.}(2019){L{\"o}hner-B{\"o}ttcher},
  {Schmidt}, {Schlichenmaier}, {Steinmetz}, \&
  {Holzwarth}}]{2019A&A...624A..57L}
{L{\"o}hner-B{\"o}ttcher}, J., {Schmidt}, W., {Schlichenmaier}, R.,
  {Steinmetz}, T., \& {Holzwarth}, R. 2019, \aap, 624, A57

\bibitem[{{L{\"o}hner-B{\"o}ttcher} {et~al.}(2018){L{\"o}hner-B{\"o}ttcher},
  {Schmidt}, {Stief}, {Steinmetz}, \& {Holzwarth}}]{2018A&A...611A...4L}
{L{\"o}hner-B{\"o}ttcher}, J., {Schmidt}, W., {Stief}, F., {Steinmetz}, T., \&
  {Holzwarth}, R. 2018, \aap, 611, A4

\bibitem[{{Meunier} {et~al.}(2010){Meunier}, {Desort}, \&
  {Lagrange}}]{2010A&A...512A..39M}
{Meunier}, N., {Desort}, M., \& {Lagrange}, A.~M. 2010, \aap, 512, A39

\bibitem[{{Neckel}(1999)}]{1999SoPh..184..421N}
{Neckel}, H. 1999, \solphys, 184, 421

\bibitem[{{Neckel} \& {Labs}(1994)}]{1994SoPh..153...91N}
{Neckel}, H. \& {Labs}, D. 1994, \solphys, 153, 91

\bibitem[{{Osipov} \& {Vasilyeva}(2019)}]{2019KPCB...35...85O}
{Osipov}, S.~N. \& {Vasilyeva}, I.~E. 2019, Kinematics and Physics of Celestial
  Bodies, 35, 85

\bibitem[{{Palumbo} {et~al.}(2022){Palumbo}, {Ford}, {Wright}, {Mahadevan},
  {Wise}, \& {L{\"o}hner-B{\"o}ttcher}}]{2022AJ....163...11P}
{Palumbo}, Michael~L., I., {Ford}, E.~B., {Wright}, J.~T., {et~al.} 2022, \aj,
  163, 11

\bibitem[{{Pereira} {et~al.}(2009{\natexlab{a}}){Pereira}, {Asplund}, \&
  {Kiselman}}]{2009A&A...508.1403P}
{Pereira}, T.~M.~D., {Asplund}, M., \& {Kiselman}, D. 2009{\natexlab{a}}, \aap,
  508, 1403

\bibitem[{{Pereira} {et~al.}(2009{\natexlab{b}}){Pereira}, {Kiselman}, \&
  {Asplund}}]{2009A&A...507..417P}
{Pereira}, T.~M.~D., {Kiselman}, D., \& {Asplund}, M. 2009{\natexlab{b}}, \aap,
  507, 417

\bibitem[{{Reiners} {et~al.}(2016{\natexlab{a}}){Reiners}, {Lemke}, {Bauer},
  {Beeck}, \& {Huke}}]{2016A&A...595A..26R}
{Reiners}, A., {Lemke}, U., {Bauer}, F., {Beeck}, B., \& {Huke}, P.
  2016{\natexlab{a}}, \aap, 595, A26

\bibitem[{{Reiners} {et~al.}(2016{\natexlab{b}}){Reiners}, {Mrotzek}, {Lemke},
  {Hinrichs}, \& {Reinsch}}]{2016A&A...587A..65R}
{Reiners}, A., {Mrotzek}, N., {Lemke}, U., {Hinrichs}, J., \& {Reinsch}, K.
  2016{\natexlab{b}}, \aap, 587, A65

\bibitem[{{Salit} {et~al.}(2004){Salit}, {Sansonetti}, {Veza}, \&
  {Travis}}]{2004JOSAB..21.1543S}
{Salit}, M.~L., {Sansonetti}, C.~J., {Veza}, D., \& {Travis}, J.~C. 2004,
  Journal of the Optical Society of America B Optical Physics, 21, 1543

\bibitem[{{Schmidt} {et~al.}(1999){Schmidt}, {Stix}, \&
  {W{\"o}hl}}]{1999A&A...346..633S}
{Schmidt}, W., {Stix}, M., \& {W{\"o}hl}, H. 1999, \aap, 346, 633

\bibitem[{Schäfer {et~al.}(2020)Schäfer, Royen, Zapke, Ellwarth, \&
  Reiners}]{10.1117/12.2560156}
Schäfer, S., Royen, K., Zapke, A.~H., Ellwarth, M., \& Reiners, A. 2020, in
  Ground-based and Airborne Instrumentation for Astronomy VIII, ed. C.~J.
  Evans, J.~J. Bryant, \& K.~Motohara, Vol. 11447, International Society for
  Optics and Photonics (SPIE), 2187 -- 2208

\bibitem[{{Shine} {et~al.}(2000){Shine}, {Simon}, \&
  {Hurlburt}}]{2000SoPh..193..313S}
{Shine}, R.~A., {Simon}, G.~W., \& {Hurlburt}, N.~E. 2000, \solphys, 193, 313

\bibitem[{{Snodgrass} \& {Ulrich}(1990)}]{1990ApJ...351..309S}
{Snodgrass}, H.~B. \& {Ulrich}, R.~K. 1990, \apj, 351, 309

\bibitem[{{Stenflo}(2015)}]{2015A&A...573A..74S}
{Stenflo}, J.~O. 2015, \aap, 573, A74

\bibitem[{{Stief} {et~al.}(2019){Stief}, {L{\"o}hner-B{\"o}ttcher}, {Schmidt},
  {Steinmetz}, \& {Holzwarth}}]{2019A&A...622A..34S}
{Stief}, F., {L{\"o}hner-B{\"o}ttcher}, J., {Schmidt}, W., {Steinmetz}, T., \&
  {Holzwarth}, R. 2019, \aap, 622, A34

\bibitem[{{Stix}(2004)}]{2004suin.book.....S}
{Stix}, M. 2004, {The sun : an introduction}

\bibitem[{{Takeda}(2022)}]{2022SoPh..297....4T}
{Takeda}, Y. 2022, \solphys, 297, 4

\bibitem[{{Takeda} \& {UeNo}(2017)}]{2017PASJ...69...46T}
{Takeda}, Y. \& {UeNo}, S. 2017, \pasj, 69, 46

\bibitem[{{Takeda} \& {UeNo}(2019)}]{2019SoPh..294...63T}
{Takeda}, Y. \& {UeNo}, S. 2019, \solphys, 294, 63

\bibitem[{{Uitenbroek} \& {Criscuoli}(2011)}]{2011ApJ...736...69U}
{Uitenbroek}, H. \& {Criscuoli}, S. 2011, \apj, 736, 69

\bibitem[{{V{\"o}gler} {et~al.}(2005){V{\"o}gler}, {Shelyag}, {Sch{\"u}ssler},
  {Cattaneo}, {Emonet}, \& {Linde}}]{2005A&A...429..335V}
{V{\"o}gler}, A., {Shelyag}, S., {Sch{\"u}ssler}, M., {et~al.} 2005, \aap, 429,
  335

\bibitem[{{Yu} {et~al.}(2014){Yu}, {Drouin}, \& {Miller}}]{2014JChPh.141q4302Y}
{Yu}, S., {Drouin}, B.~J., \& {Miller}, C.~E. 2014, \jcp, 141, 174302

\end{thebibliography}

\clearpage
\begin{appendix}
\onecolumn
        \section{Additional figures}
        \label{ap:fig}
        \begin{figure}[ht]
                \centering
                \includegraphics[width=0.49\textwidth]{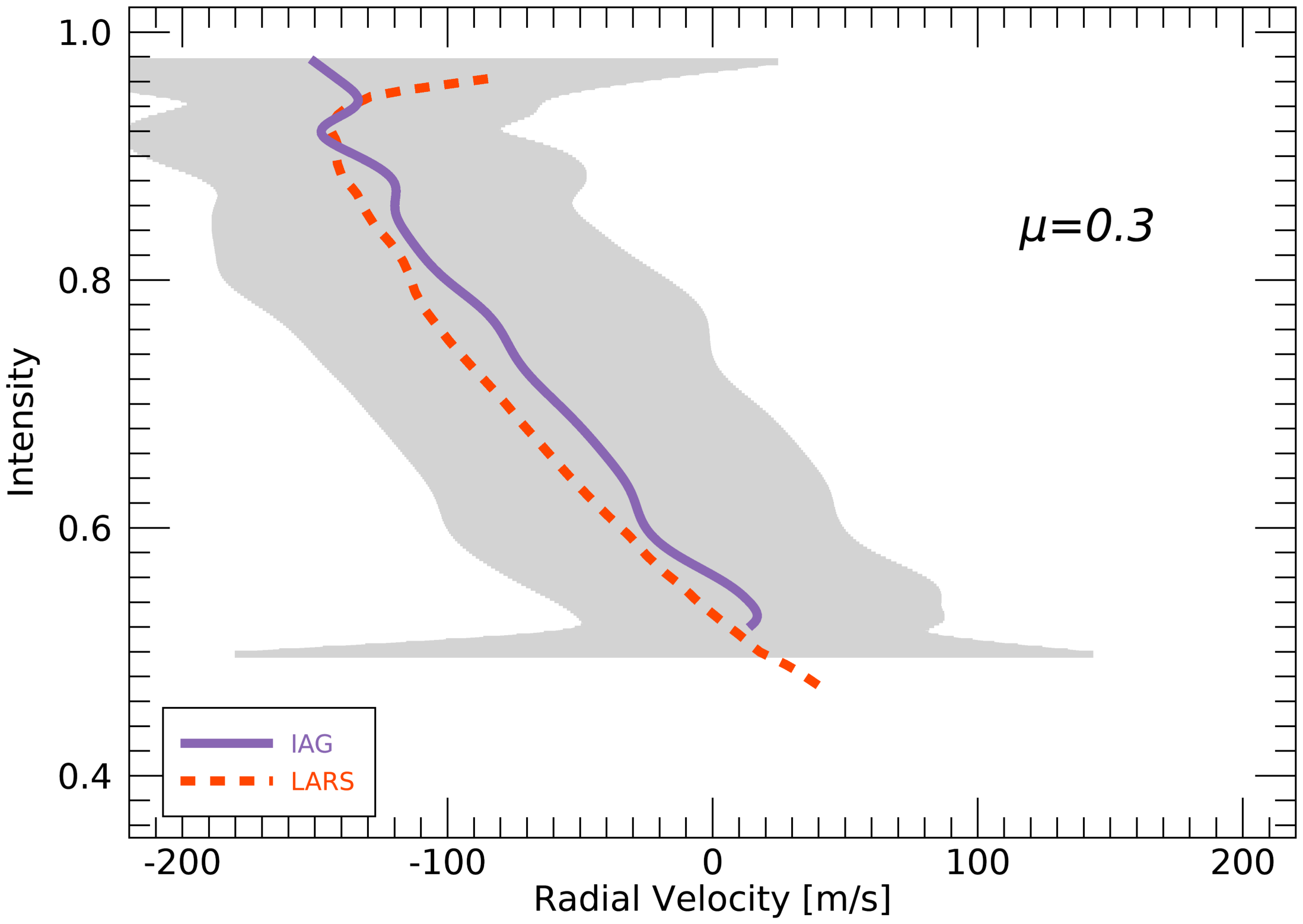}
                \includegraphics[width=0.49\textwidth]{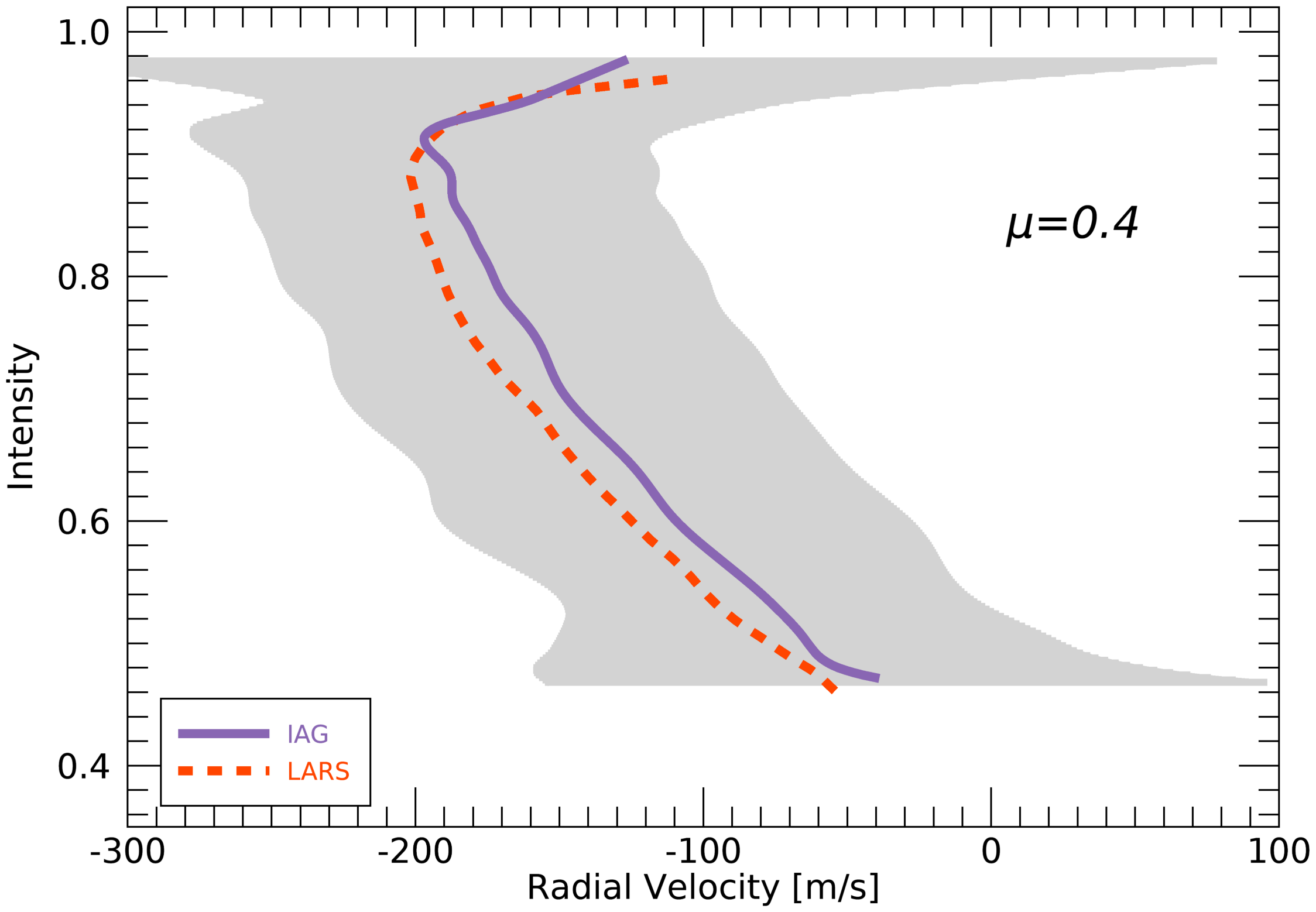}
                \includegraphics[width=0.49\textwidth]{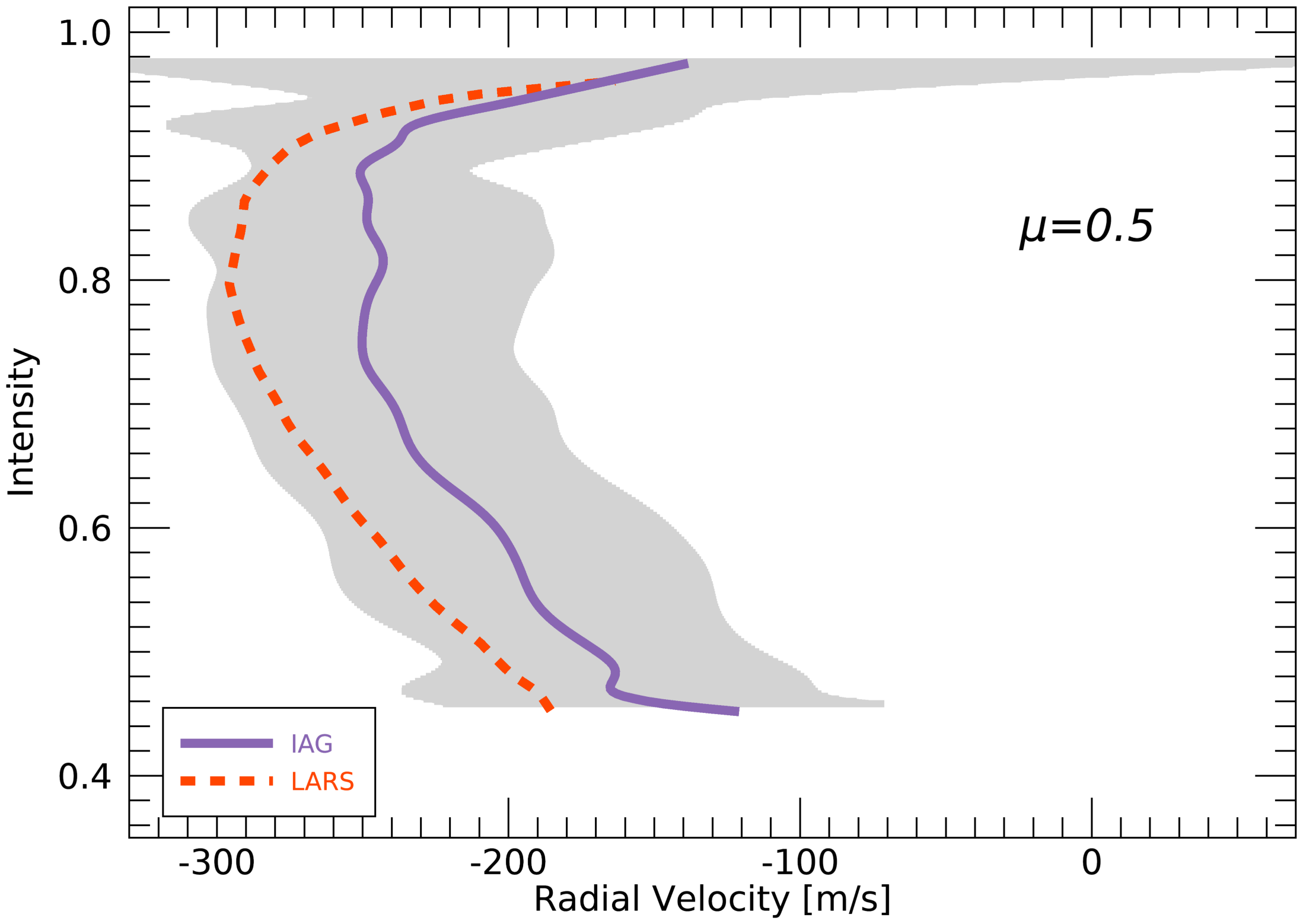}
                \includegraphics[width=0.49\textwidth]{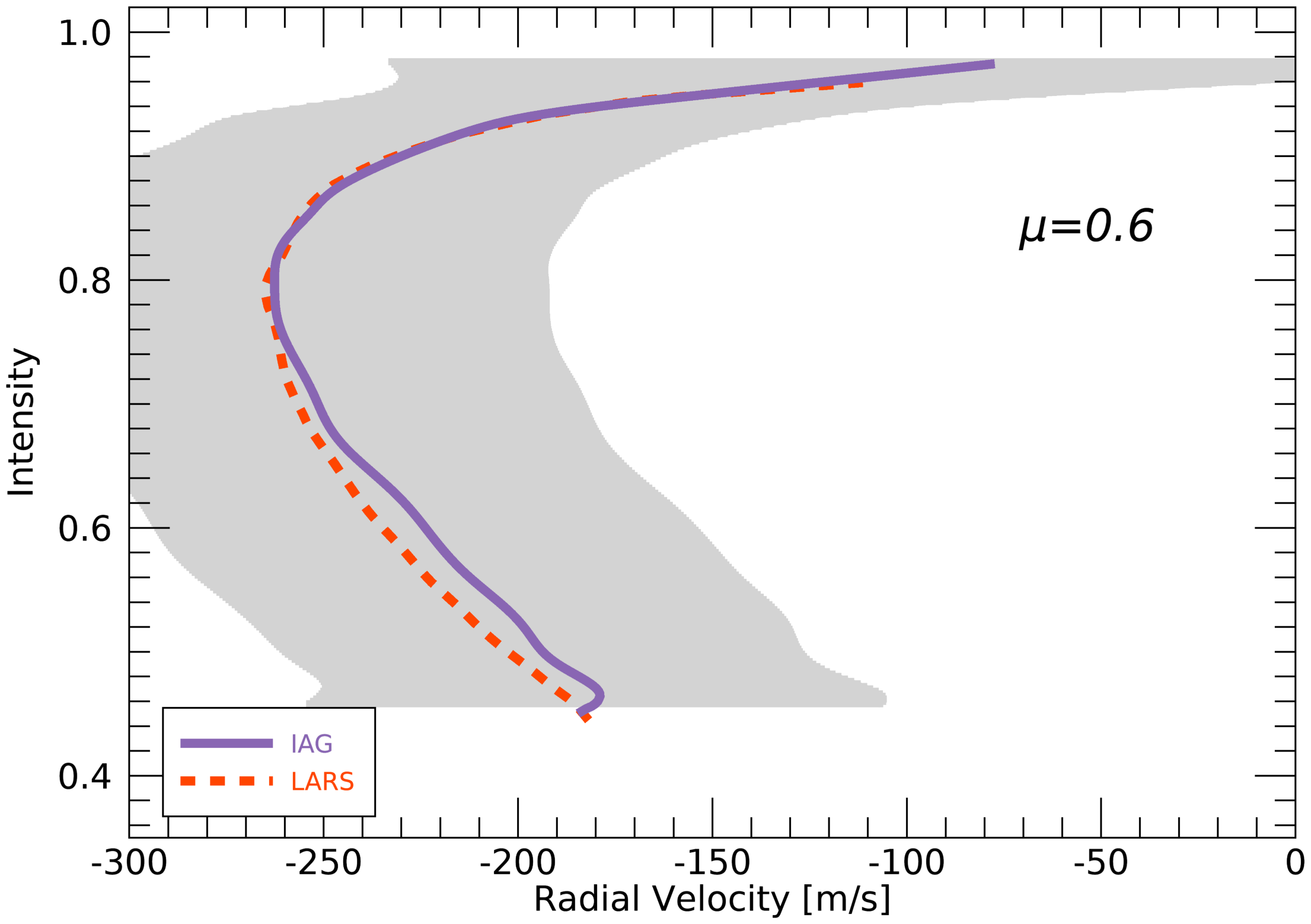}
                \includegraphics[width=0.49\textwidth]{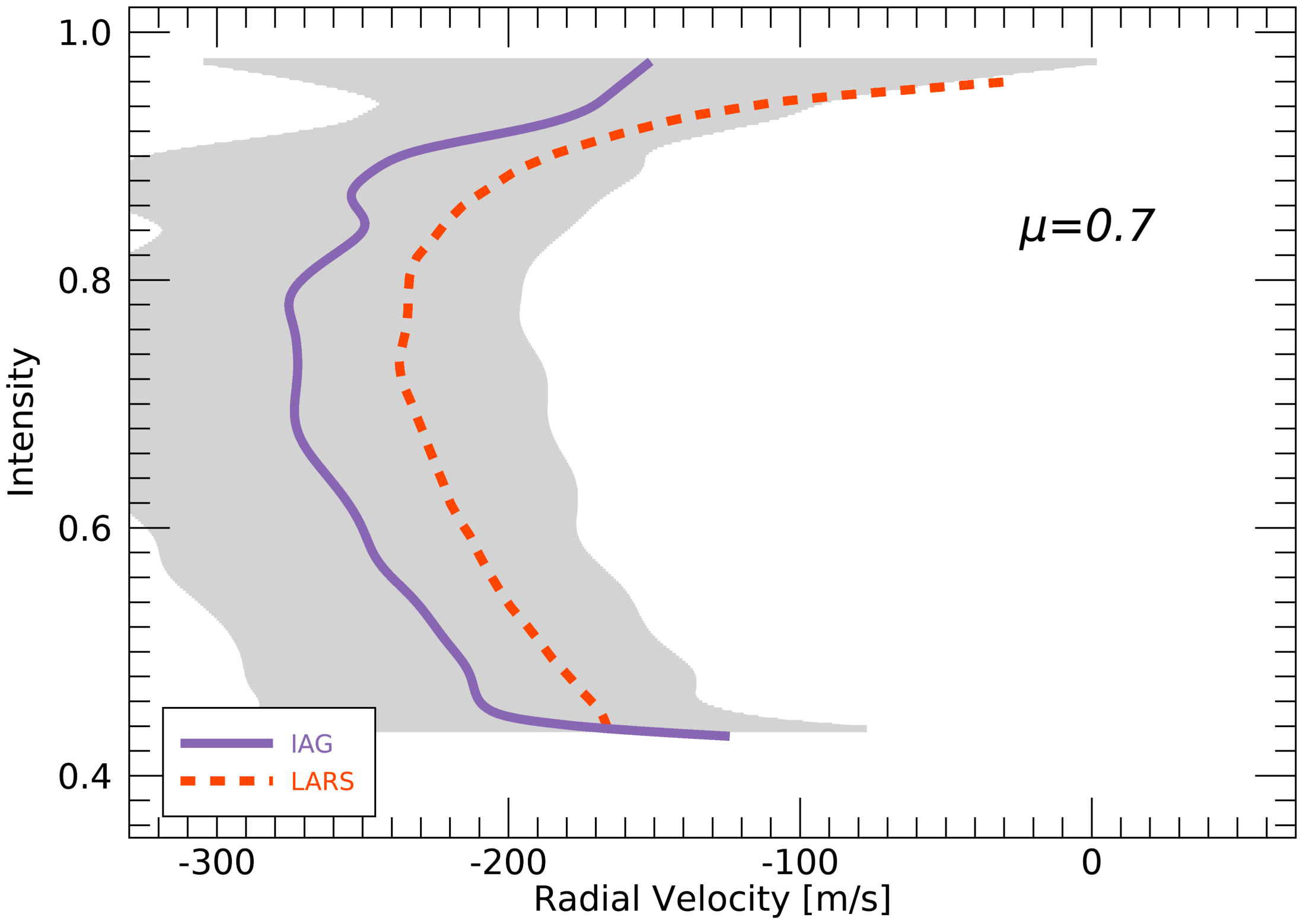}
                \includegraphics[width=0.49\textwidth]{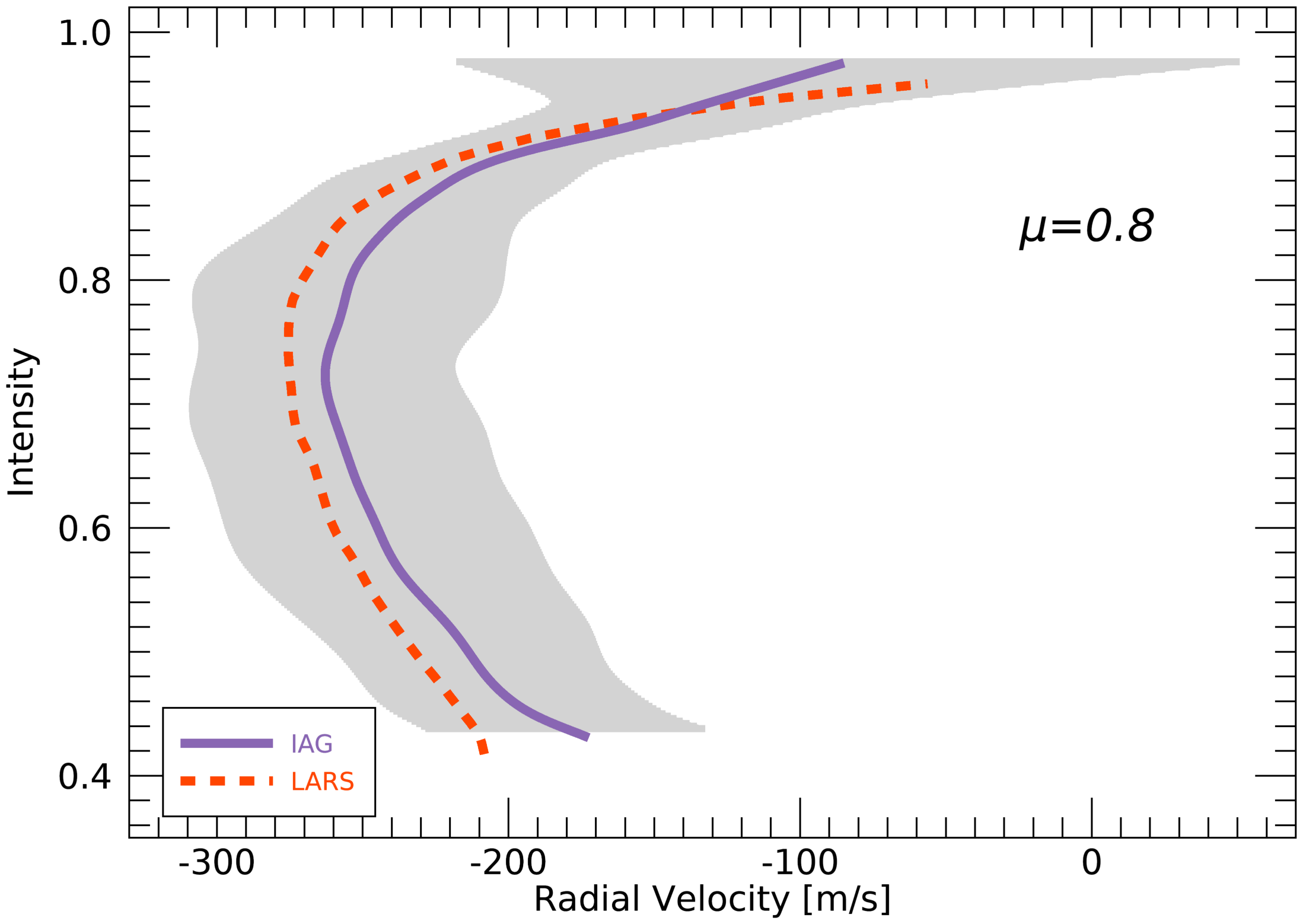}
                \caption{Bisectors of $\mu$=0.3, 0.4, 0.5, 0.6, 0.7, and 0.8 for the 6175\,\r{A} line.}
                \label{fig:mu03}
        \end{figure}
        \begin{figure}
                \centering
                \includegraphics[width=0.49\textwidth]{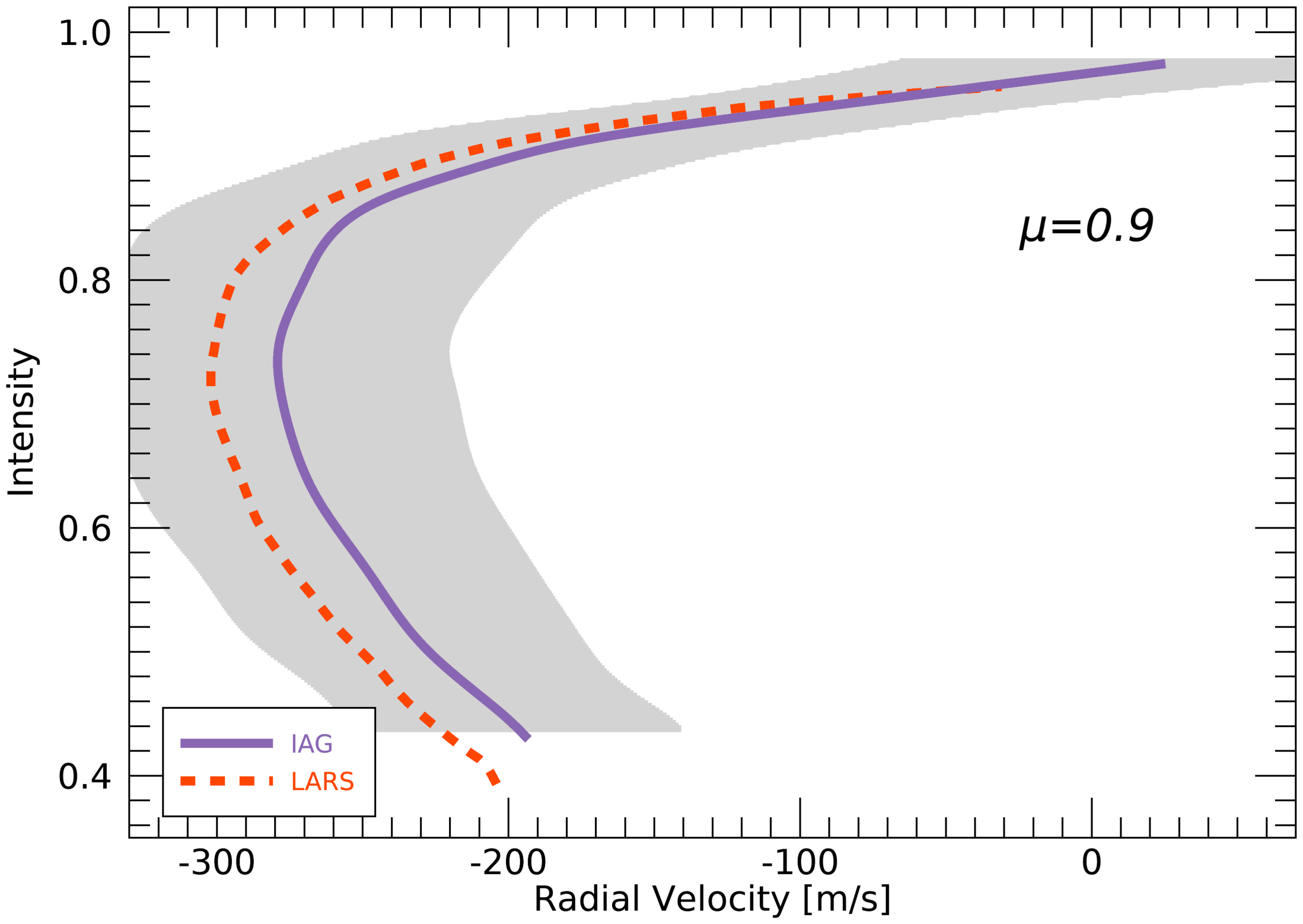}\\
                \includegraphics[width=0.49\textwidth]{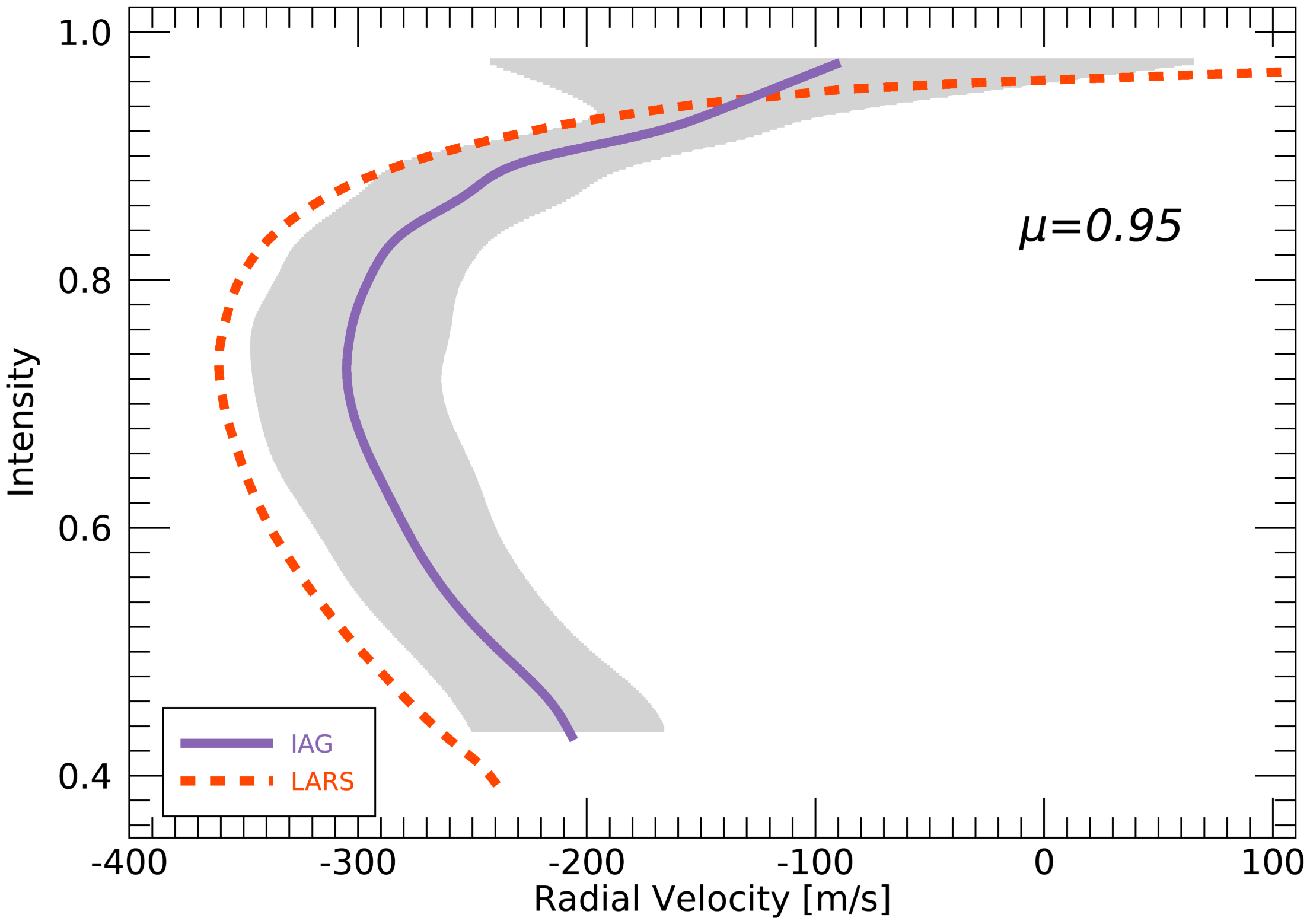}\\
                \includegraphics[width=0.49\textwidth]{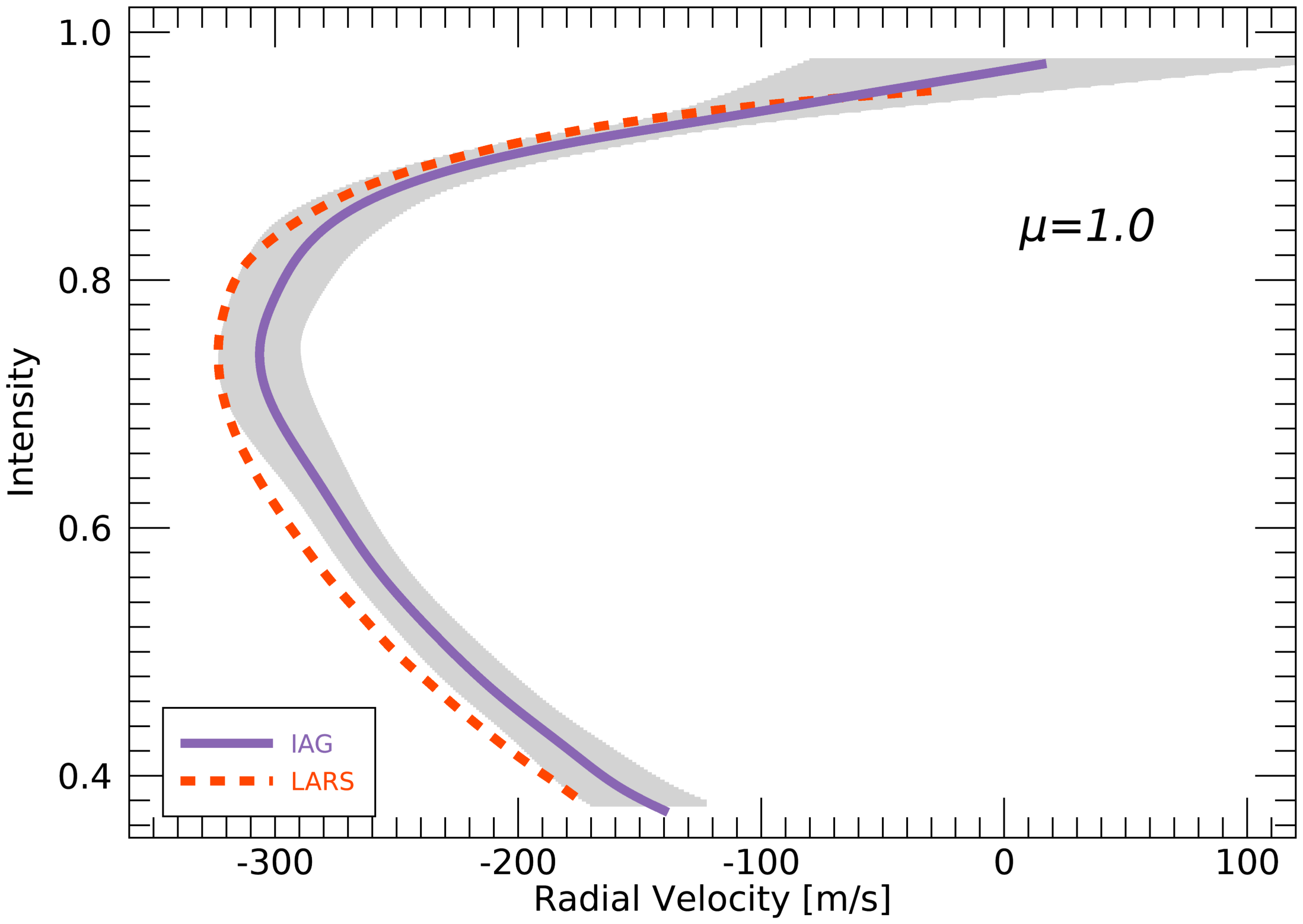}
                \caption{Bisectors of $\mu$=0.9, 0.95, and 1.0. for the 6175\,\r{A} line.}
                \label{fig:mu09}
        \end{figure}
        \twocolumn
\end{appendix}

\end{document}